\documentclass{article}
\usepackage[a4paper, total={6in,9in}]{geometry}
\usepackage[utf8]{inputenc}
\usepackage{graphicx}
\usepackage{authblk}
\usepackage[square,numbers]{natbib}
\usepackage{amsmath}

\begin{document}

\title{Multicycle terahertz pulse generation by optical rectification in LiNbO$_3$, LiTaO$_3$, and BBO crystals}

\author{Dogeun Jang and Ki-Yong Kim\textsuperscript{*}}

\affil{Institute for Research in Electronics and Applied Physics, University of Maryland, College Park, MD 20742}

\affil{*kykim@umd.edu}

\maketitle

\begin{abstract}
We report multicycle, narrowband, terahertz radiation at 14.8 THz produced by phase-matched optical rectification of femtosecond laser pulses in bulk lithium niobate (LiNbO$_3$) crystals. Our experiment and simulation show that the output terahertz energy greatly enhances when the input laser pulse is highly chirped, contrary to a common optical rectification process. We find this abnormal behavior is attributed to a linear electro-optic (or Pockels) effect, in which the laser pulse propagating in LiNbO$_3$ is modulated by the terahertz field it produces, and this in turn drives optical rectification more effectively to produce the terahertz field. This resonant cascading effect can greatly increase  terahertz conversion efficiencies when the input laser pulse is properly pre-chirped with additional third order dispersion. We also observe similar multicycle terahertz emission from lithium tantalate (LiTaO$_3$) at 14 THz and barium borate (BBO) at 7 THz, 10.6 THz, and 14.6 THz, all produced by narrowband phase-matched optical rectification.
\end{abstract}

%%%%%%%%%%%%%%%%%%%%%%%%%%  body  %%%%%%%%%%%%%%%%%%%%%%%%%%
\section{Introduction}
Intense, singlecycle, broadband terahertz (THz) sources are essential for many applications including THz-driven acceleration of electrons and protons \cite{nanni2015terahertz, palfalvi2014evanescent}, molecular alignment \cite{kampfrath2013resonant}, high harmonic generation \cite{schubert2014sub}, and material sciences \cite{nicoletti2016nonlinear}. In particular, femtosecond laser-based optical rectification (OR) in $\chi^{(2)}$ nonlinear materials is considered to be one of the most efficient methods for energy-scalable THz generation \cite{lee2009principles}. OR can be highly effective when the group velocity of the laser pulse is matched to the phase velocity of the THz wave in the nonlinear medium---called phase matching. As an OR-based THz source, lithium niobate (LN) is widely used due to its excellent material properties such as high nonlinearities ($d_{33}$ = 168 pm/V at 1THz)\cite{hebling2008generation}, high transparency at 0.4$\sim$5 $\mu$m \cite{palik1997lithium}, and well-developed poling techniques \cite{lee2009principles}. For efficient phase matching in LN, tilted-pulse-front (TPF) schemes can be used to generate intense singlecycle THz pulses  \cite{hebling2008generation, fulop2014efficient, ravi2014limitations, wu2018highly}.
%However, there has not been much practical research to develop intense multicycle narrowband THz pulse sources.

Multicycle narrowband THz sources are also of great interest owing to many emerging applications including waveguide-based electron acceleration \cite{wong2013compact}, coherent X-ray generation \cite{kartner2016axsis}, resonant pumping of materials \cite{kampfrath2013resonant}, and narrowband spectroscopy \cite{lee2009principles}.  Multicycle narrowband THz radiation is often produced by OR in periodically-poled lithium niobate (PPLN) crystals \cite{lee2000generation, l2007generation1, l2007generation2, carbajo2015efficient, jolly2019spectral}. Cryogenic cooled PPLN crystals are also used to suppress strong THz absorption in LN, lately providing a laser-to-THz conversion efficiency up to 0.1\% \cite{carbajo2015efficient}. Another approach is to drive OR with intensity-modulated laser pulses such that the produced THz waveform can follow the intensity envelope of the modulated laser pulse \cite{chen2011generation}. Other methods include transient polarization gratings \cite{stepanov2004generation}, TPF planer waveguides \cite{lin2009generation}, and cascaded second-order processes \cite{cirmi2017cascaded}. 

Recently, we have observed a new type of multicycle radiation at $\sim$15 THz emitted from a bulk LN crystal when irradiated by femotsecond laser pulses \cite{jang2019hidden}. High-energy THz radiation up to 0.7 mJ has been also produced from a large diameter (75 mm) LN wafer with 80 TW laser pumping \cite{jang2020generation}. 
%Recently, a new type of multicycle narrowband emission at 15 THz was observed from a bulk LN crystal when irradiated by femotsecond laser pulses \cite{jang2019hidden}. We have also produced high energy (up to 0.7 mJ) THz radiation by using a 80 TW pump laser along with a large diameter (75 mm) thin (35 $\mu$m) LN wafer \cite{jang2020generation}. 
%Recently, we observed a new type of multicycle radiation at 15 THz emitted from a bulk LN crystal when irradiated by femotsecond laser pulses \cite{jang2019hidden}. We have also produced high energy (up to 0.7 mJ) THz radiation by using a 80 TW pump laser along with a large diameter (75 mm) thin (35 $\mu$m) LN wafer \cite{jang2020generation}. 
This type of radiation originates from a narrow phase matching condition naturally satisfied in between two phonon resonance frequencies in LN \cite{jang2019hidden, jang2020generation}. Previously, similar narrowband radiation around 15 THz was produced by difference frequency generation (DFG) in LN, in which two separate laser pulses with different frequencies are mixed to generate THz radiation at the difference frequency  \cite{lin2012efficient}. By contrast, our THz generation method is based upon OR of a single laser pulse. %, which  by mixing of the various frequency components within the spectral bandwidth of single chirped pump pulse.
This OR process is expected to produce higher THz energy with reducing laser driver's pulse duration. 
%Generally, this optical rectification process is expected to produce higher THz energy with reducing laser driver's pulse duration. 
%Together with optical rectification, the phase matching process is expected to produce higher THz energy with reducing laser driver's pulse duration. %Together with optical rectification, this phase matching process predicts that the output THz energy will increase with decreasing laser pulse duration.
%that the more THz radiation can be generated with the shorter laser pulse duration. 
%According to optical rectification, this process is expected to produce higher THz energy with be more efficient when the pump laser pulse becomes shorter.  
However, certain LN crystals exhibit enhanced THz radiation when driven by highly chirped laser pulses \cite{jang2020generation}, contrary to our understanding of OR. Moreover, in the previous experiments, the radiation spectrum was poorly characterized with THz bandpass filter sets \cite{jang2020generation} or incompletely studied \cite{jang2019hidden}.

In this paper, we present a comprehensive study of multicycle narrowband THz generation around 15 THz from LN crystals. %by measuring  THz field autocorrelation and spectral power signals 
Experimentally, we measure THz field autocorrelation and spectral power  under various laser conditions, especially when the laser driver is chirped with third order dispersion. To explain our experimental observation, we carry out numerical calculations on THz generation and propagation in LN. We also describe experimental measurements of chirp-dependent narrowband THz generation from lithium tantalate and barium borate crystals.

%We note that similar narrowband radiation around 15 THz was previously produced by difference frequency generation (DFG) in LN \cite{lin2012efficient} in which two separate laser pulses with different frequencies are mixed to generate THz radiation at the difference frequency. %beams as the pump were obtained from a residual Nd:YAG laser beam at 1.064 $\mu$m and a frequency tunable beam from the mater oscillator/power oscillator (MOPO).
%By contrast, our THz source is based upon optical rectification of a single laser pulse. %, which  by mixing of the various frequency components within the spectral bandwidth of single chirped pump pulse.

%We show that the narrowband THz generation can be explained by optical rectification process with phase-matching achieved in a small frequency range. Furthermore, we observe enhancement of multicycle THz pulse generation at a slightly longer pump pulse by chirp due to second- (GDD) and third-order dispersion (TOD). In order to investigate our observation, one-dimensional (1-D) numerical calculations are carried out for THz generation.
%which show that multicycle THz generation efficiency is enhanced by exciting pump pulse modulation via the Pockels effect in the LN crystal when the pump pulse is highly chirped.

\section{Experimental setup}
\begin{figure}[b!]
\centering\includegraphics[width=\textwidth,height=\textheight,keepaspectratio]{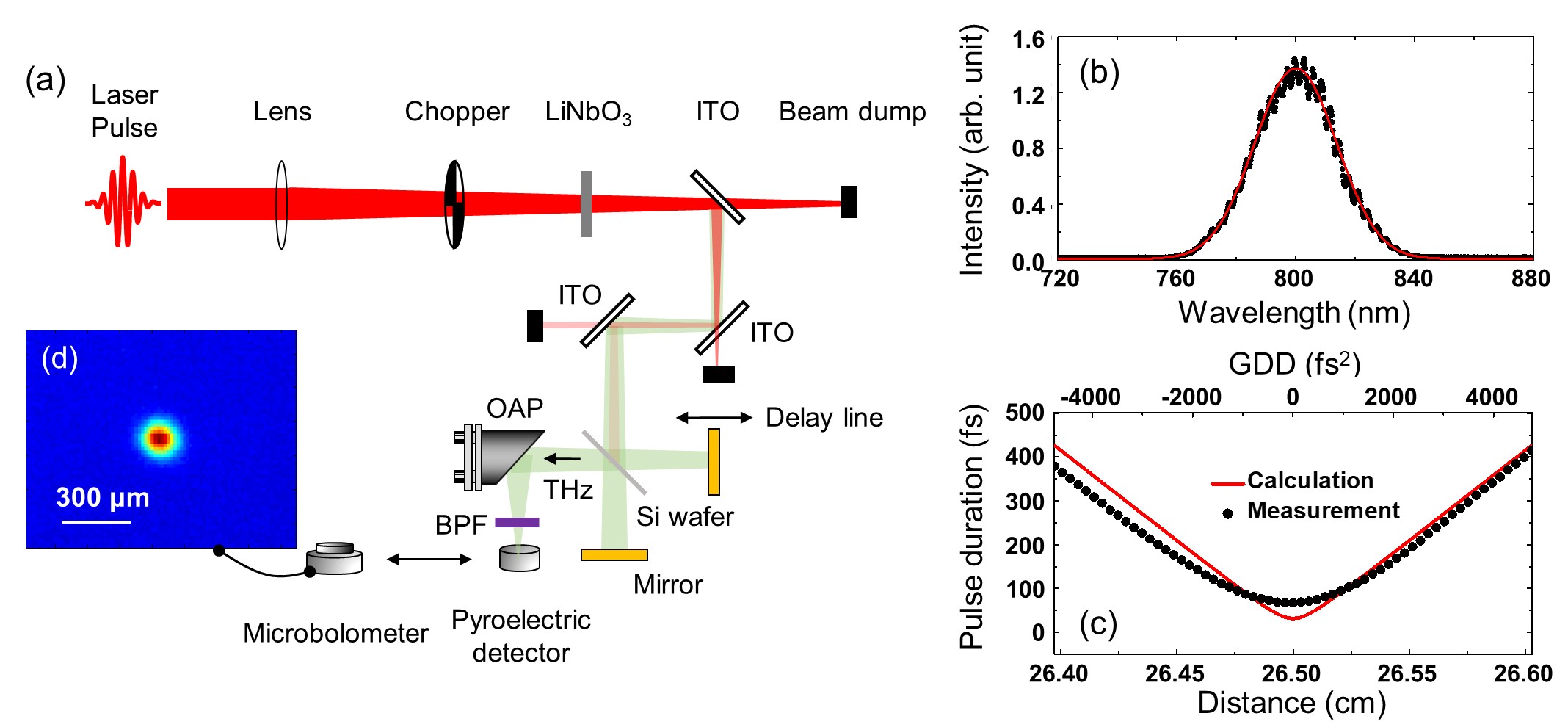}
\caption{(a) Experimental setup for multicycle THz generation and detection. (b) Input laser spectrum measured (black line) before entering the experimental setup with a Gaussian fit (red line). (c) Measured (black dotted line) and estimated (red solid line) laser pulse duration as a function of input GDD, controlled by varying the distance between the grating pair in the laser compressor. (d) Focused THz beam profile captured by a microbolometer focal plane array.}
\end{figure}

The schematic of our experimental setup is shown in Fig. 1 (a). Femtosecond laser pulses from a Ti:sapphire amplifier operating at 1 kHz are loosely focused onto a LN crystal by a lens with a focal length of 1.5 m. The laser (pump) beam size (3.4$\sim$6.8 mm in $1/e^{2}$ diameter) and fluence (3.2$\sim$28.7 mJ/cm$^{2}$) are varied by translating the LN crystal along the beam propagation direction and/or controlling the laser energy. The pump pulse provides energies up to 2.6 mJ at a central wavelength of 800 nm with a 30 nm full-width half-maximum (FWHM) bandwidth as shown in Fig. 1(b). In our measurements,  x-cut congruent LN crystals of 10 mm $\times$ 10 mm $\times$ 0.5 mm or 1 mm  (thickness) are used for THz generation. The LN crystal is oriented such that its extraordinary axis is parallel to the laser polarization for maximal THz generation.

To decouple output THz pulses from the copropagating pump beam, three optical windows coated with 180-nm-thick indium thin oxide (ITO) are placed after the LN crystal. The ITO window allows high optical transmission ($>$85\% each) in the visible and near-infrared regions with strong reflection ($\sim$80\% each) at $<$15 THz \cite{chen2010frequency}. Any pump leakage after the three ITO windows is completely blocked by a 280-$\mu$m-thick high-resistivity ($>$10 k$\Omega\cdot$cm) silicon (Si) window in the downstream beamline. %The ITO windows are used here to pre-attenuate the laser energy after pumping and thus minimize laser-induced free carrier absorption (FCA) at THz frequencies on the Si window \cite{minami2015terahertz}. 

The resulting THz pulses are characterized by a lab-built Michelson-type Fourier-transform infrared (FTIR) interferometer combined with a pyroelectric detector (PD) (Spectrum Detector Inc., API-A-62-THz). The incoming THz beam is split and recombined with a variable time delay by a 280-$\mu$m thick Si wafer in the interferometer and then focused by a 90$^{\circ}$ off-axis parabolic (OAP) mirror onto the PD detector. The THz signal from the PD detector is fed into a lock-in amplifier that is phase-locked to an optical chopper modulating the input laser beam at 10 Hz. A delay scan in the interferometer provides a THz field autocorrelation from which the spectral power can be obtained by the Fourier transform.

The pump pulse duration is varied by tuning the distance between the grating pair in the pulse compressor (see Appendix). This effectively changes the group delay dispersion (GDD) of the pump pulse. Figure 1(c) shows the pump pulse duration as a function of the grating distance. With a Gaussian spectral assumption, the Fourier transform-limited pulse duration is calculated to be  $\tau = 0.44\lambda_{0}^{2}/(c\Delta\lambda) \approx$ 31 fs, where $\lambda_{0}$ = 800 nm and $\Delta\lambda$ = 30 nm are the central wavelength and bandwidth in FWHM, respectively. With GDD = 0 fs$^{2}$, the pump pulse duration measured by a single-shot second-harmonic autocorrelator is about 67 fs in FWHM [dotted line in Fig. 1(c)]. This is longer than the transform-limited pulse duration of 31 fs. The difference is explained by uncompensated third order dispersion (TOD) and higher-order dispersion (HOD) of the pump pulse. Those result in an asymmetrical plot of the pulse duration as shown in Fig. 1(c). Figure 1(d) shows a typical THz intensity profile captured at the focus by a room-temperature microbolometer focal plane array (FLIR, Tau 2-336) \cite{yoo2016generation, jang2019spectral}. It shows that the focused THz radiation is confined within a spot size of 160 $\mu$m in FWHM diameter.

\section{Experimental results}

\subsection{Broad and narrow band THz spectrum}

\begin{figure}[b!]
\centering\includegraphics[width=\textwidth,height=\textheight,keepaspectratio]{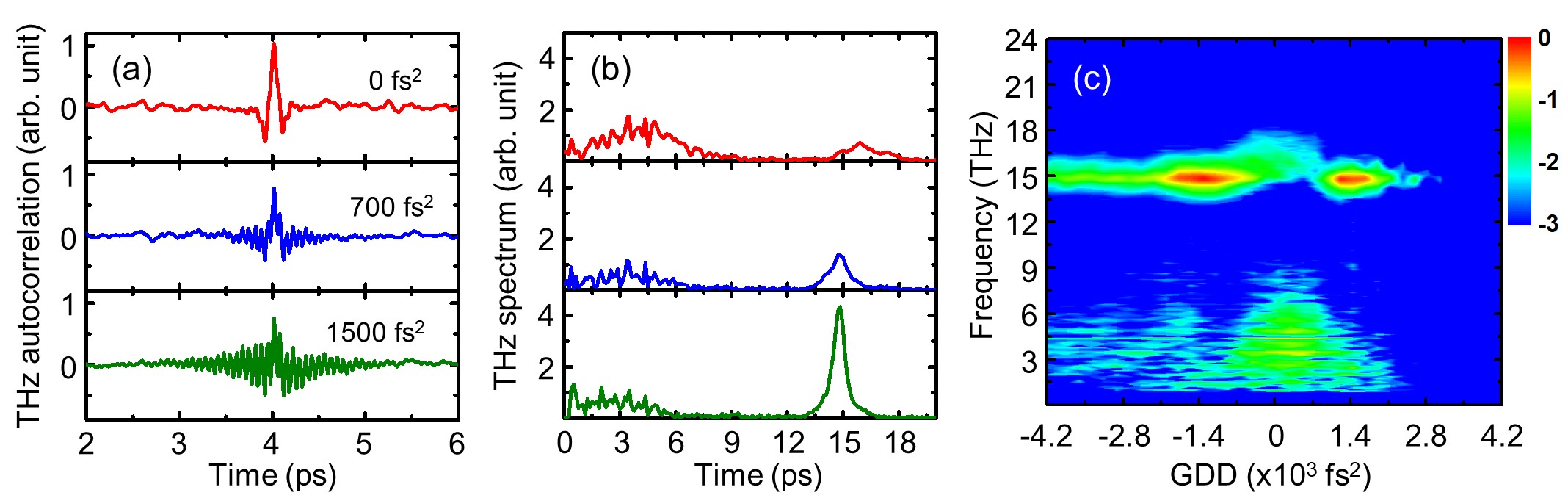}
\caption{(a) Temporal THz autocorrelation signals measured at GDD = 0 fs$^{2}$ (top), 700 fs$^{2}$ (middle), and 1,500 fs$^{2}$ (bottom). (b) Corresponding THz spectra. (c) Measured 2-D THz spectral power in a logarithmic scale (false color) as a function of the pump GDD (horizontal) and THz frequency (vertical).}
\end{figure}

Figure 2(a) shows THz autocorrelation signals obtained by the FTIR interferometer from the 1-mm-thick LN crystal with the pump fluence of 12.8 mJ/cm$^{2}$ and GDD of 0, 700, and 1,500 fs$^{2}$. The corresponding THz spectral power is obtained by the Fourier transform and plotted in Fig. 2(b). For a series of GDD values, a two-dimensional (2-D) spectral power plot (color scale) is obtained and shown in Fig. 2(c).  Figure 2 clearly shows two types of THz radiation emitted from LN---broadband singlecycle emission at 0$\sim$8 THz and narrowband multicycle emission around 15 THz. %It also shows that a transition from one to the other can be controlled by the laser GDD. 
%Both radiation are sensitive to the laser GDD. 
Interestingly, Fig. 2(c) shows that the narrowband emission is greatly enhanced with a properly stretched laser pulse duration (GDD = $-$800 fs$^{2}$, 1,500 fs$^{2}$) whereas the broadband radiation is maximally produced with the shortest pulse duration (GDD $\approx$ 0 fs$^{2}$). %We note that the GDD dependence in Fig. 2(c) appears to be unbalanced for large positive and negative GDD values. This is because the pump pulse is asymmetrically chirped as shown in Fig. 1(c). %Explain why dependence on GDD? Slight shift to the right due to material dispersion???

The broadband THz emission has been previously observed and explained by non-phase-matched OR in LN through a second-order nonlinear $\chi^{(2)}$ process \cite{l2007generation1, l2007generation2, jang2019scalable}. This yields singlecycle THz pulses emitted from both the front and rear layers of LN with a thickness of one coherence length  $l_{c} = \lambda_{\text{THz}}/(2\left|n_{g}-n_{\text{THz}}\right|)$ for each layer. Here $n_{g}$ = 2.3 is the optical group index of LN at 800 nm, $n_{\text{THz}}$ is the refractive index of LN at THz frequencies, and  $\lambda_{\text{THz}}$ is the THz wavelength. At 1 THz, $n_{\text{THz}}$ = 5.1  \cite{schall1999far, palik1997lithium} and  $\lambda_{\text{THz}}$ = 300 $\mu$m gives $l_{c}$ = 53.5 $\mu$m. This dual THz pulse generation explains the fast modulations observed in the broadband THz spectrum in Figs. 2(b) and 2(c). They arise due to interference between two temporally separated THz pulses generated from the front and rear surfaces of the LN crystal. %With the broadband THz pulse generation, the multicycle THz pulse generation can be identified by the optical rectification with the phase matching achieved in a small frequency range. 

\subsection{Characteristics of narrowband radiation at 15 THz}
%\subsection{LN thickness and energy dependence}

\begin{figure}[b!]
\centering\includegraphics[width=\textwidth,height=\textheight,keepaspectratio]{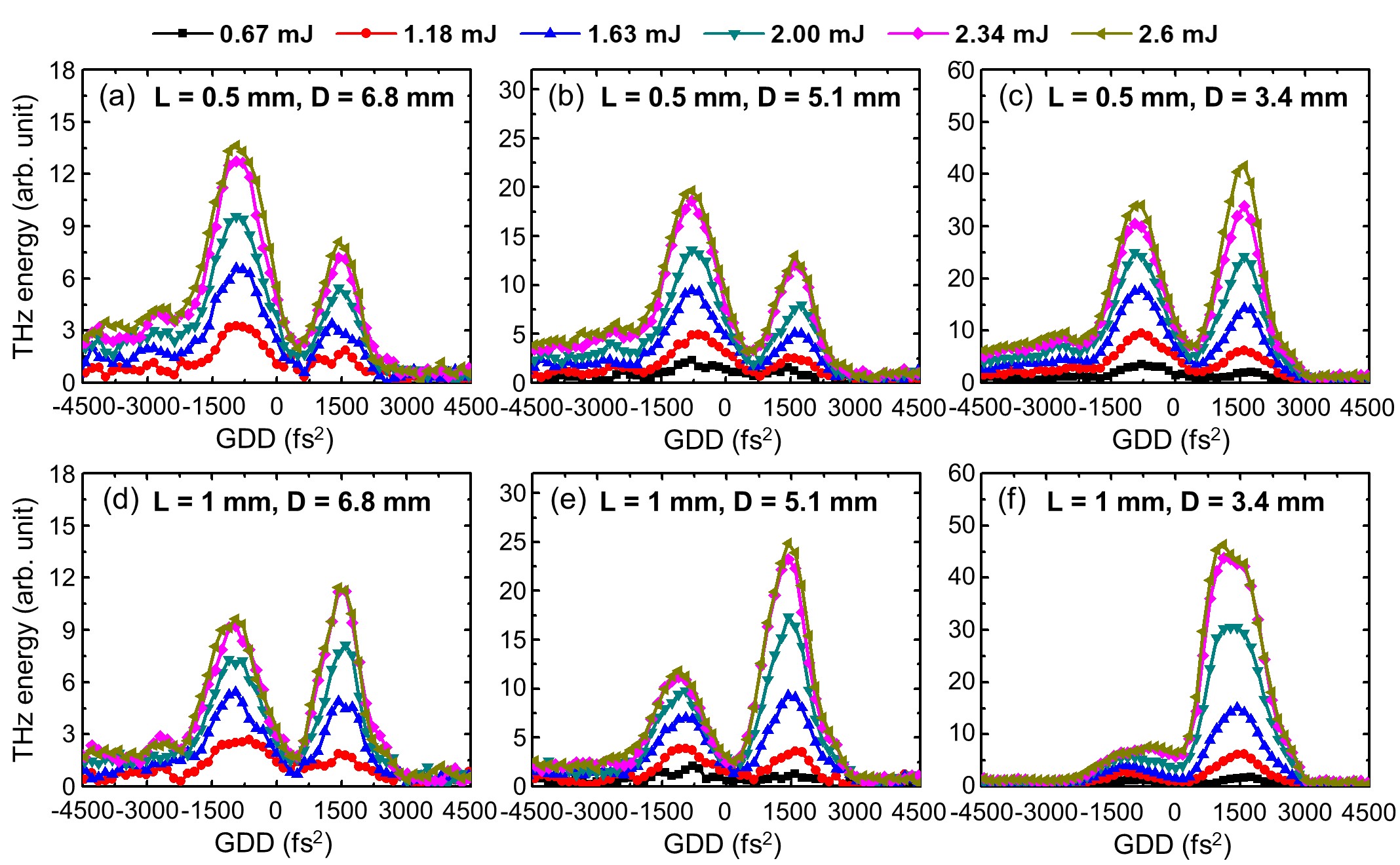}
\caption{15 THz emission energy measured as a function of the pump GDD from (a-c) 0.5-mm-thick  and (d-f) 1-mm-thick LN crystals at various pump energies and beam sizes; L is the thickness of the LN crystal, and D is the FWHM pump beam size on the crystal.} 
\end{figure}
The narrowband multicycle radiation in Fig. 2 peaks at 14.8 THz with a 0.94 THz FWHM bandwidth at GDD = 1,500 fs$^{2}$. Its output energy dependence on the pump GDD is also measured and plotted in Fig. 3. %for two different thicknesses of LN at various laser energies and beam diameters.
%A Gaussian fit to the narrowband emission in Fig. 2(a) at GDD = 1,500 fs$^{2}$ reveals narrowband radiation at 14.8 THz with a 0.94 THz FWHM bandwidth.
%Figure 3 shows the output THz energy measured as a function of the pump GDD for two different thicknesses of the LN crystal at various laser energies and beam diameters. 
Here the THz energy is measured by the pyroelectric detector (PD) along with a THz bandpass filter (BPF) providing a central frequency of 15 THz placed in front. This allows to measure only narrowband THz radiation around 14.8 THz. %Here, one of the FTIR arms is blocked not to measure any constructive/deconstructive interference.

As shown in Fig. 3, the output THz energy peaks at two GDD ranges of 900$\sim$1,600 fs$^{2}$ (positive chirp) and $-$1,200$\sim-$800 fs$^{2}$ (negative chirp). More interestingly, the positive GDD range yields more output THz energy with increasing pump energy and/or LN thickness. Furthermore, the output THz energy is abnormally suppressed at GDD $\approx$ 0 fs$^{2}$, where the highest nonlinearity is expected due to the shortest pump pulse duration. Previously, similar results were observed and explained by THz screening and absorption by free charge carriers produced by multi-photon laser absorption in LN  \cite{wu2018highly}. In our experiment, the nonlinear absorption coefficient by free carrier absorption (FCA), $\alpha_{fc}$, is estimated to 31.63 cm$^{-1}$ at 14.8 THz under laser conditions of 470 GW/cm$^{2}$ peak intensity and 800 nm wavelength \cite{zhong2015optimization, fulop2010design}. The value, however, is much smaller than the intrinsic absorption coefficient $\alpha$ = 1,440 cm$^{-1}$ at 14.8 THz in LN \cite{jang2019hidden, jang2020generation}. Therefore, three-photon-absorption followed by FCA is not believed to cause the suppressed THz emission at GDD $\approx$ 0 fs$^{2}$. Instead, the odd GDD-dependence can be explained by a THz-induced cascaded effect on the pump pulse that has nonzero TOD as will be described in Section 5.

Another interesting feature observed with the narrowband THz radiation is its energy scaling. Figure 4(a) shows the measured output THz energy emitted from the 1-mm-thick LN crystal as a function of the pump energy with the beam diameter of D = 3.4 mm, corresponding to Fig. 3(f). %Here the optimal GDD value for maximal THz generation shifts with increasing pump energy as shown in Fig. 3, and the GDD values in Fig. 4(a) are taken from Fig. 3 at the positive optimal GDD values for each pump energy.
In phase-matched OR, the output THz energy is expected to increase quadratically with the pump intensity, i.e., $e_{\text{THz}} \propto \left| I_{\text{pump}} \right|^{2} $. This is indicated by the red-dotted line in Fig. 4(a). The observed THz energy, however, increases much faster than the theoretical prediction at the pump energy exceeding 1.2 mJ. This unexpected behavior can be also explained by a THz-induced cascaded effect as will be explained in Section 5.

The resulting laser-to-THz conversion efficiency is shown in Fig. 4(b). The maximum THz output energy of $\sim$92.6 nJ and efficiency of $3.7 \times 10^{-5}$ are achieved. This provides a maximum field strength of 0.4 MV/cm at the focus, estimated from the measured energy, pulse duration ($\sim$470 fs), and beam spot size ($\sim$160 $\mu$m). %\textbf{The conversion efficiency is lower than the value obtained with a thin () LN in our previous report \cite{jang2020generation}. What's pump fluence?  More details are described in Section XXX.}

\begin{figure}[t!]
\centering\includegraphics[width=\textwidth,height=\textheight,keepaspectratio]{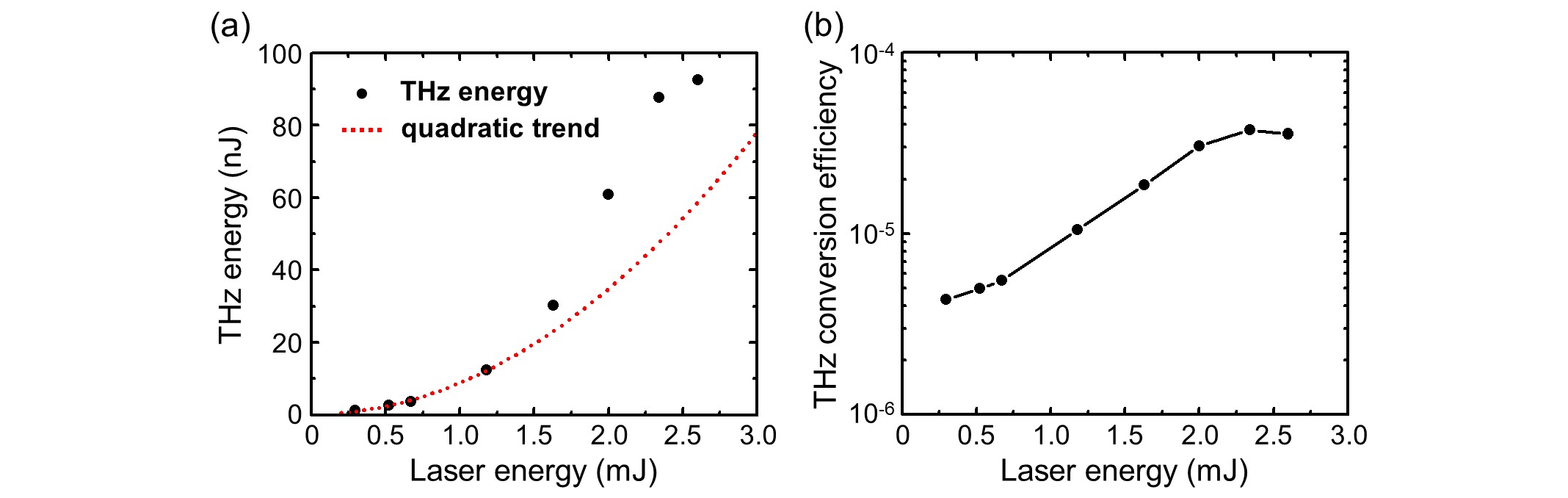}
\caption{(a) Measured multicycle THz pulse energy as a function of the pump energy (black dots) obtained with L = 1 mm and D = 3.4 mm in Fig. 3(f), coplotted with a quadratic trend line (red dotted line) expected from OR. (b) Corresponding laser-to-THz conversion efficiency.}
\end{figure}

\subsection{Narrowband THz radiation from LT and $\beta$-BBO crystals}

\begin{figure}[t!]
\centering\includegraphics[width=\textwidth,height=\textheight,keepaspectratio]{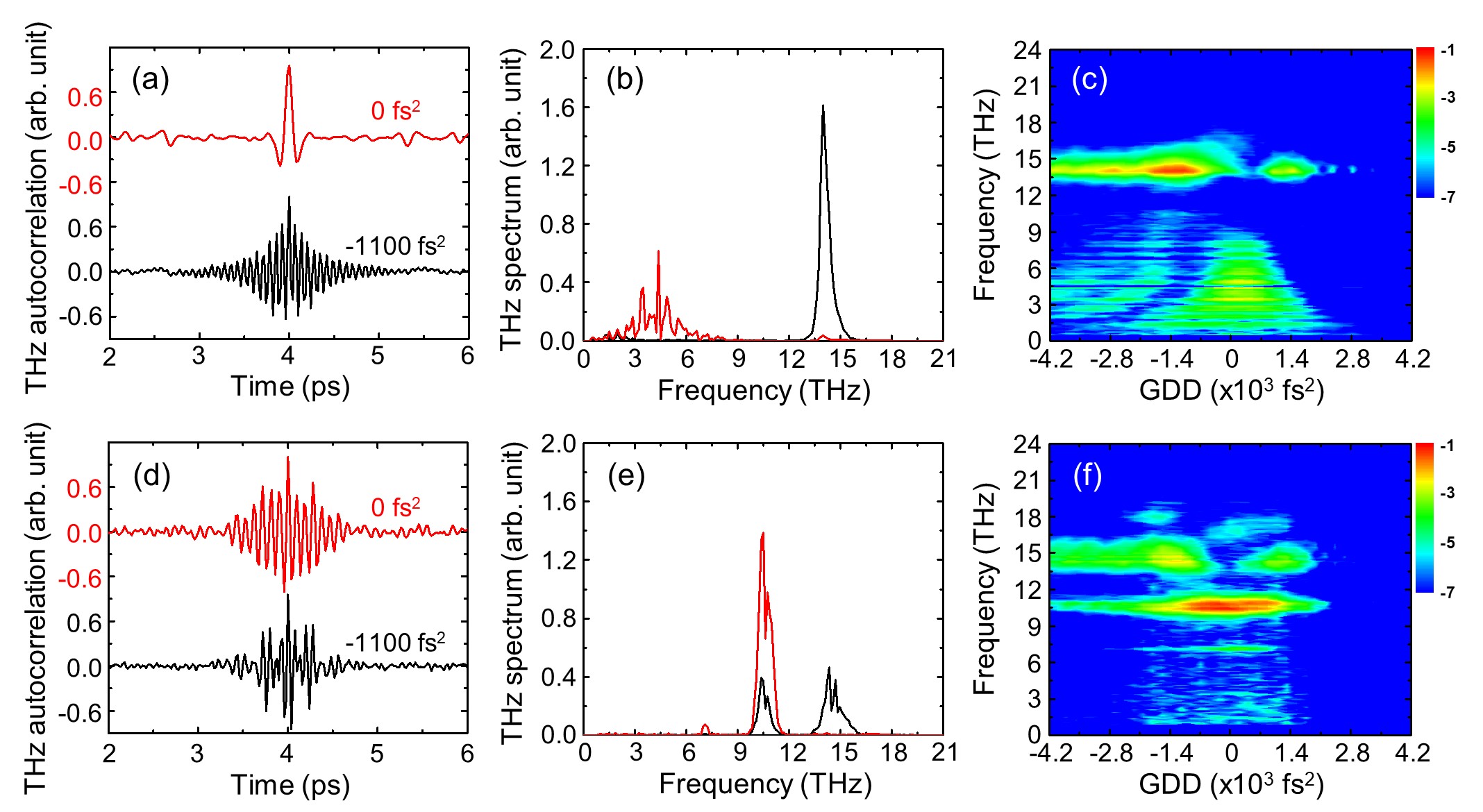}
\caption{(a) Temporal THz autocorrelation signals measured from a 0.5-mm-thick LT crystal at GDD = 0 fs$^2$ (top red line) and $-$1,100 fs$^2$ (bottom black line). (b) Fourier-transformed THz spectra. (c) 2-D plot of THz spectrum (false color) measured as a function of the pump GDD. (d-f) Data measured with a 0.1-mm-thick $\beta$-BBO crystal.
%Temporal THz autocorrelation signals measured from a 0.5-mm-thick LT crystal in (a) and a 0.1-mm-thick $\beta$-BBO crystal in (d) at GDD = 0 fs$^{2}$ and -1,100 fs$^{2}$. Corresponding THz spectra in (b) and (e). Measured 2-D spectral power in (c) and (f) as a function of the pump GDD (horizontal) and THz frequency (vertical).
}
\end{figure}

Multicycle narrowband THz emission is also observed from lithium tantalate (LT) and beta-barium borate ($\beta$-BBO). These two nonlinear materials including LN are commonly used inorganic $\chi^{(2)}$ nonlinear crystals and have a trigonal structure with point group 3m. LT and $\beta$-BBO are also tested for GDD-dependent THz generation, and the result is shown in Fig. 5.  %In addition to LN crystals, chirp-dependent multicycle THz generation is also tested with other commonly used inorganic $\chi^{(2)}$ nonlinear materials---lithium tantalate (LT) and beta-barium borate ($\beta$-BBO).
Figures 5(a) shows THz autocorrelation signals obtained from a 0.5-mm-thick LT crystal at two different pump GDD values of 0 and $-$1,100 fs$^{2}$. The corresponding spectra are shown in Fig. 5(b). From a GDD scan from $-$4,200 fs$^2$ and 4,200  fs$^2$,  a 2-D plot of THz spectrum (color scale) is obtained and displayed in Fig. 5(c). Figures 5(d-f) shows experimental data obtained with a 0.1-mm-thick $\beta$-BBO crystal.
%It shows output THz spectral power maps obtained from (a) a 0.5-mm-thick LT crystal and (b) a 0.1-mm-thick $\beta$-BBO crystal measured as a function of the pump GDD. Figures 5(b) and 5(d) are their spectral line-outs at two GDD values ($-$1,100 fs$^{2}$ and 0 fs$^{2}$). 
For both measurements, the pump energy fluence is fixed at 28.7 mJ/cm$^{2}$.  Clearly, both crystals exhibit multicycle THz waveforms and consequent narrowband THz emission. In the case of LT, its narrowband emission is centered at 14 THz, consistent with the refractive index \cite{gervais1997lithium} and phase matching condition for LT. Similar to the LN crystals tested before, LT shows both singlecycle (broadband) and multicycle (narrowband) radiation depending on the pump chirp condition.
%$\beta$-BBO crystals are commonly used for phase-matched second harmonic generation (SHG) at 800 nm wavelength

The spectral power 2-D plot (color scale) of $\beta$-BBO shown in Fig. 5(f) exhibits narrowband emission at 7 THz, 10.6 THz, and 14.6 THz. Our result is consistent with a previous study reporting narrowband emission at 4.3 THz, 7 THz, and 10.6 THz \cite{valverde2017multi}. Interestingly, 4.3-THz emission is not seen in our experiment possibly due to its weak spectral power. Instead, our experiment reveals new narrowband emission at 14.6 THz. %To the best of our knowledge, these are the first reports of the multicycle THz emission in two nonlinear materials.
We note that this observation was possible due to our FTIR-based detector's capability of measuring high-frequency THz emission beyond 10 THz. Contrary to commonly used electro-optic sampling (EOS) methods, in our scheme the detection bandwidth is not limited by the laser pulse duration or THz absorption/dispersion in the electro-optic (EO) material. Also, our detector is independent from the source and not affected by any pump chirp. This allows us to characterize the radiation spectrum without being distorted or restricted by the pump GDD. 

%In the $\beta$-BBO crystal, the different behaviors of the multicycle THz generations between 14.6 THz and lower frequencies were observed in Fig. 9(c). This indicates that the pump pulse length corresponding to at least two THz field variations is required for efficient multicycle THz pulse generation by intensity-modulated pump pulse. At GDD = -1,100 fs$^{2}$, the effective pump length (full-width $1/e^{2}$) is estimated to 160 fs, which longer than 2.2 times oscillation cycle of the THz field at 14 THz. But, for a lower frequency radiation, the effective pump length can be achieved at higher GDD values. However, it leads to decreasing pump intensity, and weakening the THz generation as shown in Fig. 9(c).

\section{Theoretical background}

\begin{figure}[t!]
\centering\includegraphics[width=\textwidth,height=\textheight,keepaspectratio]{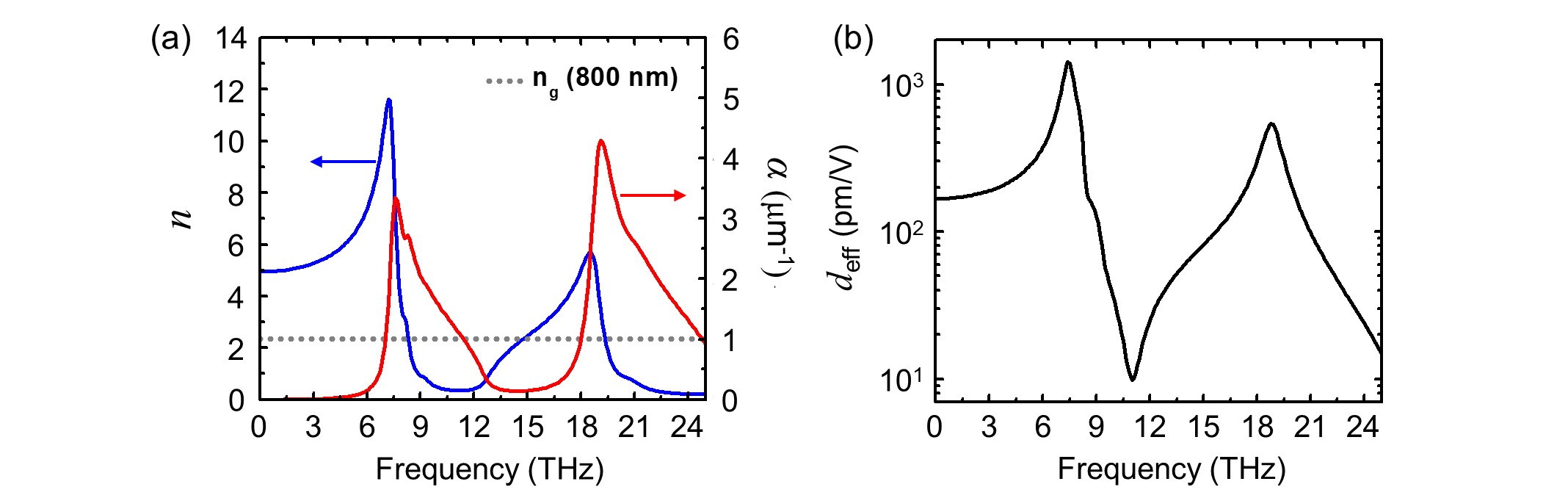}
\caption{(a) Refractive index (blue solid line) and absorption coefficient (red solid line) of congruent LN in the extraordinary direction, co-plotted with the group index at 800 nm (gray dotted line). (b) Calculated effective nonlinear coefficient $d_{\text{eff}}$ of LN in a logarithmic scale as a function of THz frequency.}
\end{figure}

%The multicycle THz generations at these frequencies are very interesting discovery, the detailed physical explanation of those multicycle THz generations however are not described in this paper due to the lack of completed study. The similar behaviors of multicycle THz pulse generations from three different nonlinear materials are observed at frequencies $\geq$ 14 THz. 

The narrowband THz radiation observed in Figs. 2 and 5 is fundamentally characterized by phase-matched ($n_{g} = n_{\text{THz}}$) OR. Here $n_{g}$ and $n_{\text{THz}}$ are the group and refractive indices at the pump and THz frequencies, respectively. For example, Fig. 6(a) shows the refractive index $n_{\text{THz}}$ of congruent LN as a function of frequency \cite{palik1997lithium, ding2011efficient, sussman1970tunable, barker1967dielectric} (see Appendix). The optical group index $n_{g}$ = 2.3 at 800 nm is also plotted in Fig. 3(a) with a gray dotted line. It shows that the phase-matching condition, $n_{g}$ = $n_{\text{THz}}$, is satisfied at 14.8 THz in between two strong transverse-optical (TO) phonon resonance frequencies in LN (7.4 THz and 18.8 THz). These resonance frequencies are clearly shown by the absorption coefficient $\alpha$ plotted in Fig. 3(a) with a red solid line. Note that there are two additional phase-matched frequencies occurring at 8.3 THz and 19.3 THz. However, little or no emission is expected at those frequencies because of their strong absorption in LN.

Figure 6(b) shows the effective nonlinear coefficient $d_{\text{eff}}$ in the extraordinary direction of congruent LN \cite{ding2011efficient, sussman1970tunable}. It shows $d_{\text{eff}}$ reaches its local minimum value of 10 pm/V at 11 THz while peaking to 1,424 pm/V and 543 pm/V at the two phonon resonance frequencies, 7.4 and 18.8 THz, respectively. At frequencies $<$7.4 THz, $d_{\text{eff}}$ asymptotically approaches 168 pm/V, which is consistent with the reported value at 1 THz in Ref. \cite{hebling2008generation}. At 14.8 THz,  $d_{\text{eff}}$ = 82 pm/V, which is still sufficiently large to generate strong THz radiation \cite{hebling2008generation, fulop2010design}.

We emphasize this type of narrow phase matching can occur in many nonlinear crystals including LT and BBO. In general, the refractive index $n_{\text{THz}}$ changes so large in between two phonon resonance, and often there exist one or multiple frequencies at which $n_{\text{THz}}$ becomes equal to the optical group index of refraction $n_{g}$. %In the crystals, the refractive index $n_{\text{THz}}$ changes greatly in between two phonon resonance and can be momentarily matched to the optical group index of refraction $n_{g}$.
Absorption is also expected to be relatively low in between phonon resonance frequencies. In addition, the phase-matched frequency can be tuned by varying the pump laser wavelength although it provides a narrow tuning range. For example, optical pumping at 0.4$\sim$1.9 $\mu$m in LN can yield phase-matched emission at 14.4$\sim$15.7 THz.

%In general, the thickness of the LN crystal should be compatible to the effective length $L_{\text{eff}}$ for efficient THz generation \cite{ding2011efficient}. At 14.8 THz, the absorption coefficient reaches its minimum value of $\alpha$ = 1,440 cm$^{-1}$ which however is sufficiently large, i.e., $\alpha^{-1} \approx 7 \mu\text{m} \ll L$, where the $L$ is the length of the LN crystal. This would indicate that a layer with length $L_{\text{eff}}$ = $\alpha^{-1}$ in the front surface of the LN crystal can efficiently contribute the THz emission.

At the phase-matched frequency of 14.8 THz in LN, the absorption coefficient reaches its minimum value of $\alpha$ = 1,440 cm$^{-1}$ as shown in Fig. 6(a). This value, however, is still large enough to attenuate the emission significantly. In phase-matched OR, the effective length for maximal THz generation is generally given by $L_{\text{eff}} = {(\alpha/2 - \alpha_L)}^{-1}\ln{[\alpha / (2 \alpha_L)] }$ $\approx$ 160 $\mu$m, where $\alpha_{L}$= 0.0078 $\text{cm}^{-1}$ is the laser absorption coefficient near 800 nm \cite{jang2020generation, hattori2007simulation}. This means that only a layer of 160 $\mu$m thickness from the front surface of the LN crystal can maximally generate 14.8 THz radiation when the incident pump pulse is unchirped. In this case, a thin LN crystal is best suited for efficient THz generation as demonstrated in our previous experiment \cite{jang2020generation}. For thicker ($\gg L_{\text{eff}}$) crystals as in the current experiment, negatively-chirped pump pulses are generally preferable as they can be compressed with propagation and effectively produce THz radiation from near the rear surface of the LN crystal. 

Interestingly, our experiment shows that both positive and negative chirps yield enhanced THz emission as shown in Figs. 2 and 3. So it is the stretched pulse duration, not chirp, that matters in our narrowband THz generation. In general, too long pulses are not good as they do not provide enough bandwidths to generate 14.8 THz radiation by OR. %However, if the pulse duration is too long, then it is not suitable as it may not provide an enough bandwidth to generate radiation at 14.8 THz by optical rectification.
For example, at GDD = 1,600 fs$^{2}$ that yields enhanced 14.8 THz radiation, the corresponding pump pulse duration is estimated to $\sim$130 fs in FWHM from Fig. 1(c). This is about twice longer than the period of 14.8 THz radiation, 68 fs. This implies that the pump pulse must have certain intensity modulations (or pulse splitting) within the pulse envelope to provide a sufficient bandwidth to generate 14.8 THz. Such modulations can be initially made by applying a proper combination of GDD and TOD onto the pump pulse.

Pump intensity modulations can also arise and be amplified from a nonlinear process. For instance, a pump pulse propagating through LN can be distorted by the THz field it produces via a linear electro-optic (or Pockels) effect \cite{shen2007nonlinear, jewariya2009enhancement}. This is a second-order $\chi^{(2)}$ nonlinear process and can lead to spectral shifts, broadening, and modulations of the pump pulse \cite{ravi2014limitations, shen2007nonlinear, hattori2007simulation}. Thus, in order to explain the peculiar GDD dependence of the narrowband radiation, one needs to include the Pockels effect, as well as laser-THz dispersion and absorption
in conducting numerical simulations.

%However, the behavior of the THz generation as shown in Fig. 2 would not be explained. The possible mechanism of the behavior in Fig. 2 could be attributed to the variation of intensity-modulated pump pulse, produced by cross-phase modulation (XPM) through the $\chi^{(2)}$ nonlinear process such as the Pockels effect in the LN crystal \cite{shen2007nonlinear,jewariya2009enhancement}. Second-order nonlinear XPM is the phase shift, which can be produced by the interaction with the electric field of another wave in the $\chi^{(2)}$ nonlinear crystal. In optical rectification, the THz field is generated by the optical pump pulse, and consequently, this can distort the pump pulse, leads to spectral shift, broadening, modulation of the pump pulse \cite{ravi2014limitations, shen2007nonlinear, lee1999nonlinear, hattori2007simulation, shen2008spatiotemporal}. The efficiency of THz generation is influenced by the variation of the pulse shape. Therefore, the variation of the pump pulse modulation due to the XPM through the Pockels effect leads to the behavior of the THz energy curves as shown in Fig. 2.

\section{Numerical simulations}
\subsection{Theoretical model}
In order to simulate THz generation by OR, we first present the electric field of a laser (pump) pulse by using a Gaussian envelope shape given by 
\begin{equation}
E\left(t\right) = E_{0}\exp{\left[-\frac{\Delta\omega^{2}t^{2}}{8\ln{2}} -j\omega_{0}t \right]},
\end{equation}
where $\omega_{0}$ is the center frequency, $\Delta\omega = 4\ln{2}/\Delta t$  is the FWHM spectral bandwidth, and $\Delta t$ is the pulse duration in FWHM. The amplitude $E_{0}$ is determined by the pump fluence $F$ as 
\begin{equation}
E_{0} = \left(\frac{\ln{2}}{\pi}\right)^{1/4}\sqrt{\frac{\Delta\omega F}{c\epsilon_{0}n_{0}}}.
\end{equation}To account for chirped pump pulses, the spectral phase of the pump pulse is expanded in a Taylor series about $\omega_{0}$ as 
%To account for high-order chirped pump pulse to optical rectification, a Taylor series of the dispersion as a function of $\omega$ about $\omega_{0}$ is given by 

\begin{equation}
\phi\left(\omega\right)=\frac{\text{GDD}}{2}\left(\omega-\omega_{0}\right)^{2} + \frac{\text{TOD}}{6}\left(\omega-\omega_{0}\right)^{3},
\end{equation}
where the fourth and higher-order terms are neglected for the sake of simplicity. The input pump pulse is obtained in the frequency domain as

\begin{equation}
E_{p}\left(\omega\right)=FT\{ E\left(t\right) \}\exp\left[-j\phi\left(\omega\right)\right],
\end{equation}
where $FT$ denotes the Fourier transform.

Then we solve one-dimensional (1-D) coupled forward Maxwell equations (FME)
%To explore the Pockels effect on the high-order chirped pump pulse for multicycle THz pulse generation in the LN crystal, an effective 1-D system of coupled Forward Maxwell Equations (FME)
\cite{husakou2001supercontinuum, jang2019scalable, jang2020generation} self-consistently for both THz and optical pump pulses  in the frequency domain as
\begin{equation}
\begin{split}
\frac{\partial E_{T}\left(\omega_{T}, \xi \right)}{\partial\xi}  = & -\left(\frac{\alpha_{T}}{2} + jD\left(\omega_{T}\right) \right)E_{T}\left(\omega_{T}, \xi \right), \\
& -j\frac{\omega_{T}\chi^{(2)}}{2cn_{0}\left(\omega_{T} \right)}\int\limits_{0}^{\infty} E_{p}\left(\omega + \omega_{T}, \xi \right)E_{p}^{*}\left(\omega, \xi \right) d\omega, 
\end{split}
\end{equation}

\begin{equation}
\begin{split}
\frac{\partial E_{P}\left(\omega, \xi \right)}{\partial\xi}  = & -\left(\frac{\alpha_{P}}{2} + jD\left(\omega\right) \right)E_{P}\left(\omega, \xi \right), \\
& -j\frac{\omega\chi^{(2)}}{2cn_{0}\left(\omega \right)}\int\limits_{0}^{\infty} E_{P}\left(\omega - \omega_{T}, \xi \right)E_{T}^{*}\left(\omega_{T}, \xi \right) d\omega_{T}, \\
& -j\frac{3\omega\chi^{(3)}}{8cn_{0}\left(\omega \right)}FT\{\left| E_{P}\left(t,\xi \right) \right|^{2}E_{P}\left(t, \xi \right) \},
\end{split}
\end{equation}
where $E_{T}$ and $E_{P}$ are the electric fields of THz and optical pump, respectively, which propagate in the coordinate $\xi = z - ct/n_{g}$ moving at the group velocity of the pump. The first terms on the right hand side of both equations ($\alpha$ terms) correspond to absorption of both fields. The second terms correspond to material dispersion $D\left(\omega_{T}, \omega \right) = \omega\left(\omega_{T}, \omega \right) \left[ n\left(\omega_{T}, \omega \right) - n_{g} \right]/c$. The third term in Eq. (5) represents the second-order nonlinear polarization due to OR, a source term for THz radiation. %, which indicates that various frequency components within the bandwidth of the pump pulse lead to the THz generation at $\omega_{T}$.
The third term in Eq. (6) describes the Pockels effect on the pump pulse induced by the produced THz field. The last term in Eq. (6) corresponds to self-phase modulation (SPM) of the pump pulse via the Kerr effect, where the third-order nonlinear susceptibility $\chi^{(3)}$ is derived from the nonlinear refractive index $n_{2} = 3\chi^{(3)}/\left( 4c\epsilon_{0}n_{0}^{2} \right)$. Here $n_{2} = 10^{-6}$ cm$^{2}$/GW is used in our calculation \cite{desalvo1996infrared}.  In the simulation, we ignore THz-induced pump modulation via the $\chi^{(3)}$ nonlinear process. This is because the $\chi^{(3)}$-based pump phase shift, $\Delta\varphi^{(3)}$, is much smaller than $\Delta\varphi^{(2)}$ induced by the Pockels effect, i.e., $\Delta\varphi^{(3)}/\Delta\varphi^{(2)} = 3\chi^{(3)}E_{\text{THz}}/(2\chi^{(2)}) \ll 1$ \cite{shen2007nonlinear}.

The numerical integrals in Eqs. (5) and (6) are solved by a 4th-order Runge-Kutta method with spatial resolution of 500 nm in order to achieve required numerical convergences. Note that the model used here considers only 1-D space along the propagating direction $z$. This is justified for a large pump beam size, where any transverse beam effects such as self-focusing and diffraction can be ignored.

In the simulation, the input pump pulse is assumed to be Gaussian with a 30-nm FWHM spectral bandwidth at 800 nm to be consistent with our experimental condition.  Experimentally, nonzero third order dispersion (TOD) arises from two sources; one is from the compressor's tuning for GDD control (see Appendix). %When tuning distance between the grating pair, not only GDD but also TOD are varied simultaneously.
The other one comes from the amplifier itself but not properly compensated by the compressor. This  residual TOD is implicitly shown in Fig. 1(c) by the asymmetric slope. Here the total TOD can be expressed as TOD = $\text{TOD}_{g} + \text{TOD}_{i}$, where the subscripts $g$ and $i$ denote ``grating" and ``initial (residual)". In the simulation, we used $\text{TOD}_{g} \approx -2(\text{fs}) \cdot\text{GDD}$ and $\text{TOD}_{i}$ = 3,800 fs$^{3}$ for our compressor system (see Appendix).

\subsection{Simulation results and discussions}
\subsubsection{GDD-dependent THz spectrum}

\begin{figure}[b!]
\centering\includegraphics[width=\textwidth,height=\textheight,keepaspectratio]{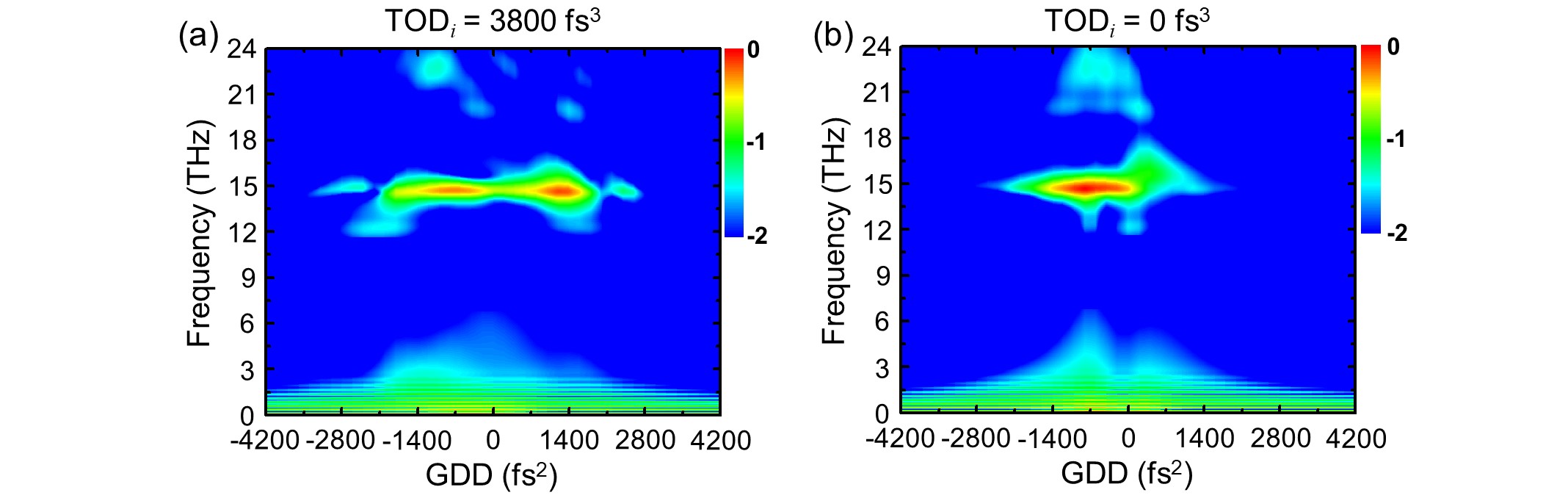}
\caption{Simulated THz spectrum (false color) in a logarithmic scale as a function of the pump GDD (horizontal axis) obtained with (a) TOD = $-2(\text{fs})\cdot\text{GDD}$ + 3,800 fs$^{3}$ and (b) TOD = $-2(\text{fs})\cdot\text{GDD}$.
}
\end{figure}

We compare our simulation results with the experimental ones shown in Fig. 2. Figure 7(a) shows a simulated THz spectral power plot (color scale) obtained from a 1-mm-thick LN at laser fluence of 13.6 mJ/cm$^{2}$. %Note the pump TOD varies as TOD = 3,800 fs$^{3}$ + $2(\text{fs})\cdot\text{GDD}$.
Clearly, it reproduces multicycle narrowband THz radiation around 15 THz. Also, it is most efficiently produced at GDD = 1,600 fs$^{2}$ and $-$800 fs$^{2}$. At GDD $\approx$ 0 fs$^2$, it is greatly suppressed while the broadband radiation at 0$\sim$8 THz is maximally enhanced. This is in good agreement with our experimental results. %verifying our simulation model for nonlinear THz generation processes in LN. %The difference in results from experiments can be explained by small uncertainty in parameters of the material and pump laser. 

For comparison, we repeated the simulation with a chirped pump with $\text{TOD}_{i}$ = 0 fs$^{3}$ to better understand the role of TOD in THz generation. %to show the THz generation driven by the linearly chirped pulses, i.e., TOD = 0 fs3, in conjunction with the Pockels effect as shown in Fig. 7(b).
The result is shown in Fig. 7(b). Interestingly, it shows the narrowband emission singly peaks at GDD = $-$800 fs$^{2}$, and the maximal THz energy increases by 18.9\% compared to Fig. 7(a). %We can see in Fig. 7(b) that an optimal GDD value for multicycle THz generation is GDD = -800 fs$^{2}$ for 1 mm thick LN crystal and 18.9\% enhancement of THz energy is observed compare to a result of maximum THz energy in Fig. 5(a).
This does not agree with our experiment but is consistent with a general OR process, in which the radiated THz energy increases with decreasing pump pulse duration at fixed laser energy. %In the general principle of optical rectification process, the radiated THz energy strongly increase when the pump pulse duration is reduced in the undepleted pump approximation.
Here a negatively chirped pump pulse is more favorable for THz generation because it can be compressed as it propagates through LN that possesses positive material dispersion. %At highly negative or positive GDD values, decreasing intensity of the pump pulse results in decreasing THz energies. 
%In general, we tested it using our numerical calculations, indicating that the behavior as shown in Fig. 5(b) are mostly attributed to the variation of the pump intensity for different chirped pump pulse, since the pump pulse is modulated by the THz field.
In addition, we repeated the simulation without including the Pockels effect but keeping the original TOD---the result is not shown in Fig. 7. In this case, maximal THz radiation occurs at GDD = $-$400 fs$^{2}$, but the generated THz energy decreases to 62\% compared to the first simulation result shown in Fig. 7(a).  %simulated considered only TOD effect on the multicycle THz pulse generation but without including the Pockels effect (not shown here). In this case, the efficient THz generation was observed at GDD = -400 fs$^{2}$, but generated THz energy is decreased to 62\% of that when we considered both TOD and the Pockels effect.
All these suggest that the strange GDD dependence of 15 THz observed in our experiment can be attributed to a combined action of both the Pockels and TOD effects. %is strongly evidence that the conversion efficiency of optical rectified multicycle THz pulse increased by the modulated pump pulse via the Pockels effect induced by radiated THz field. 

\subsubsection{Evolution of THz and laser fields with propagation}

\begin{figure}[b!]
\centering\includegraphics[width=\textwidth,height=\textheight,keepaspectratio]{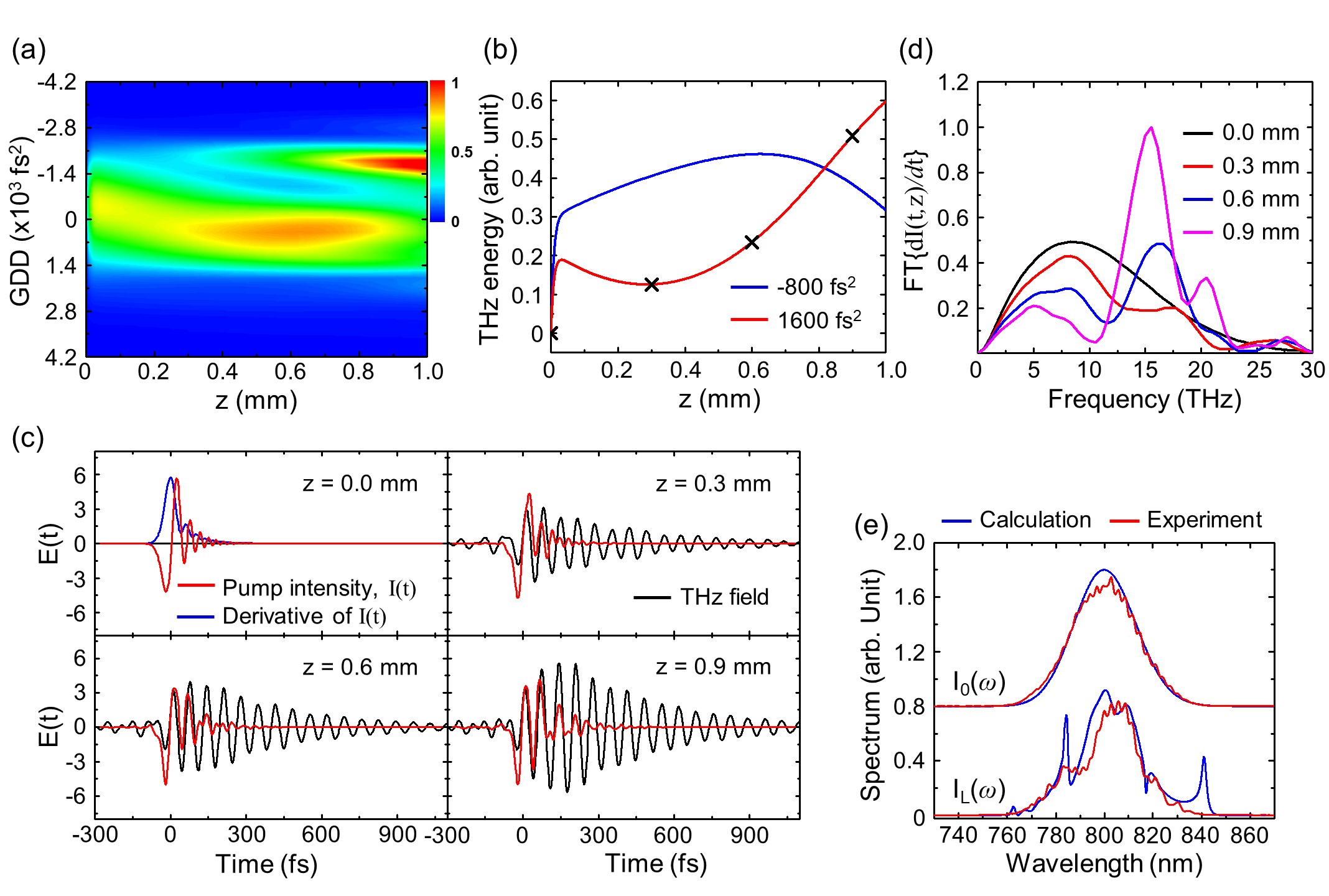}
\caption{(a) Calculated multicycle THz energy (color scale) along the propagation direction $z$ in LN for various initial GDD values. (b) Line-outs  from (a) at GDD = $-$800 fs$^2$ (blue line) and 1,600 fs$^2$ (red line). (c) THz waveforms (black lines) and pump intensity derivatives, $-dI/dt$, (red lines) computed at $z$ = 0, 0.3, 0.6, and 0.9 mm (marked "X" in (b) with GDD = 1,600 fs$^2$). The input pump intensity profile $I(t)$ is shown at $z$ = 0 mm (blue line). (d) Fourier-transformed spectra of the derivatives of pump intensity envelopes in (c). (e) Measured (red line) and simulated (blue line) pump spectra. $I_{0}\left(\omega\right)$ and $I_{L}\left(\omega\right)$ are the initial and final ($z$ = 1 mm) spectra.}
\end{figure}

For a detailed understanding of multicycle THz generation, we simulated the evolution of THz electric fields and laser intensity envelopes as they propagate through LN. %In order to detail the multicycle THz generations, the simulated variation of multicycle THz generation are shown in Fig. 8. 
Here all simulation parameters are the same as in Fig. 7(a). First, Figure 8(a) shows the cumulative THz energy (color scale) plotted as a function of the initial GDD (vertical axis) and the propagation distance $z$ (horizontal axis). Here the pump TOD varies as TOD =  $-2(\text{fs})\cdot\text{GDD}$ + 3,800 fs$^{3}$. Two line-outs at GDD = $-$800 fs$^{2}$ and 1,600 fs$^{2}$ are plotted in Fig. 8(b).  In the case of GDD = $-$800 fs$^{2}$, the generated THz energy slowly increases and then decreases beyond $z$ = 0.6 mm. However, with GDD = 1,600 fs$^{2}$, the energy noticeably increases after $z $ = 0.3 mm.

Figure 8(c) shows the THz electric fields (black lines) calculated at $z$ = 0, 0.3, 0.6, and 0.9 mm with an initial GDD value of 1,600 fs$^{2}$. Also co-plotted (red lines) are the temporal derivatives of the pump intensity profiles, $-dI\left( t, z \right)/dt$. For reference, the input pump intensity profile $I(t)$ is shown (blue line) at $z$ = 0 mm. Clearly, it exhibits damped intensity modulations on its tail due to nonzero GDD and TOD.
%One can see that the GDD and TOD chirped pump pulse resembles a squared and damped Airy function \cite{cui2016spectral}, 
Initially, this type of intensity modulations is not best suited for multicycle THz generation because its oscillation frequency is chirped and not fully matched to 14.8 THz. However, with a propagation, the pump envelope becomes synchronously modulated by the co-propagating 14.8 THz field via the Pockels effect. This results in a series of pump pulses (pulse splitting) separated by the THz period, which in turn drives OR resonantly to generate multicycle 14.8 THz radiation. %with propagation the pump envelope gets modulated by the co-propagating THz field via the Pockels effect. This leads to variation between the pump intensity and the THz field in the time domain. Therefore, this nonlinear modulation process eventually results in enhanced multicycle THz radiation. 
This is evidently shown by the time derivative of the pump intensity envelope in Fig. 8(c) (red lines). Its Fourier spectral power at various $z$ is plotted in Fig. 8(d). At $z$ = 0.9 mm, it peaks at $\sim$15 THz. It also produces two side bands. The left one is responsible for singlecycle broadband THz radiation at $<$10 THz, whereas the right (weak) one is believed to be the source of  20$\sim$23 THz shown in Fig. 7(a) although it was not observed in our experiment.
%The difference behavior in Fig. 6(b) for GDD = -800 fs$^{2}$ and 1600 fs$^{2}$ could be explained by variation of the dispersion. The dispersion comes from the chirp of the pump pulse and the LN crystal. They have different signs and values, and the combination of these can change the intensity and the shape of the pump pulse through the pump propagation in the LN crystal. Therefore, the competition between the dispersion and XPM leads to different behavior for negative and positive GDD. 

Also, the input $I_{0}\left(\omega \right)$ and transmitted $I_{t}\left(\omega \right)$  pump spectra are computed and plotted in Fig. 8(e) along with experimentally measured ones.  The transmitted one corresponds to a 1-mm-thick LN crystal pumped at 22 mJ/cm$^{2}$ with GDD = 1,500 fs$^{2}$. As shown in Fig. 8(e), both simulated and measured pump spectra show spectral modulations and small frequency shifts. For negative GDD values, blue-shifted spectra are observed (not shown here). 
%The spectrum distortion of the pump is induced by the Pockels effect due to singlecycle and multicycle THz pulses as shown in Fig. 2, respectively. In general, a frequency shift can occur when the pump pulse is shorter than the THz one  \cite{shen2007nonlinear}. In our experiment, this occurs at GDD = 0 fs$^{2}$. The singlecycle THz pulse duration is about 470 fs in FWHM estimated from the autocorrelation signal for GDD = 0 fs$^{2}$ as shown in Fig. 2(b).  At GDD = 1,600 fs$^{2}$, the laser pulse duration is about 130 fs in FWHM. In contrast, the spectral modulation is occurred when the pump pulse is comparable to the variation of the field, which is about 68 fs for generated multicycle THz pulse. Therefore, the observed spectrum in Fig. 6(e) is directly resulted in the Pockels effect.

Figure 9 shows the  simulated output energy of multicycle THz radiation as a function of the pump GDD for 0.5-mm and 1-mm thick LN crystals. %The frequency filtering between 10 THz and 20 THz is applied to resulting calculation to ignore contributions of other frequency components.
It shows that more output THz energy is produced at negative pump GDD values with a thinner (0.5 mm) LN. With increasing laser fluence and LN thickness (1.0 mm), however, the peak moves to the positive GDD side, consistent with our measurement in Fig. 3. Also, our simulation provides a laser-to-THz conversion efficiency of $2.1 \times 10^{-5}$ from a 1-mm-thick LN pumped at 13.6 mJ/cm$^2$. This is in reasonable agreement with our experimental value  $\sim$10$^{-5}$ obtained under similar laser fluence conditions in Fig. 4(b).
%It shows that  output energy is produced with the the positively chirped pump pulse is more suitable for multicycle THz pulse generation when pump energy/fluence are increased. In addition, more output energy is produced from a 1-mm-thick LN than 0.5 mm thick, which is consistent with our measurement in Fig. 3. We obtained the conversion efficiency of $2.1 \times 10^{-5}$ in this calculation, which is twice higher than the measurement due to the different excitation of the pump fluence, but still in good agreement with the experimental results. The conversion efficiency is measured to be $\sim 10^{-5}$. 

\begin{figure}[h!]
\centering\includegraphics[width=\textwidth,height=\textheight,keepaspectratio]{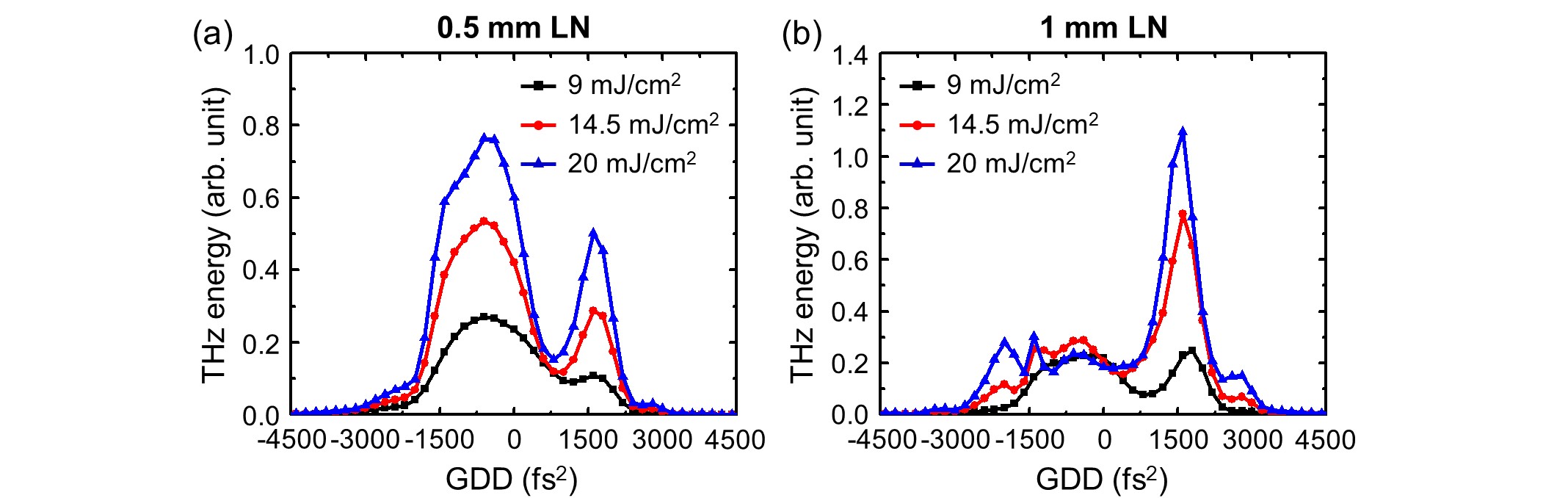}
\caption{Simulated output energy of multicycle THz radiation as a function of the pump GDD for (a) 0.5-mm-thick and (b) 1-mm-thick LN crystals at three pump fluences.}
\end{figure}

\section{Conclusion}
In conclusion, we have demonstrated efficient multicycle narrowband THz generation at 14.8 THz from bulk LN crystals by using chirped optical pump pulses. The generation mechanism is explained by phase-matched OR naturally occurring in between two phonon resonance frequencies in LN. In our experiment, we have observed enhanced multicycle THz emission when the pump pulse is highly chirped. This anomalous behavior is also observed in our numerical simulations and explained by resonant intensity modulations of the pump pulse by self-produced THz fields through the Pockels effect. The modulated pump pulse can in turn produce multicycle THz radiation efficiently with propagation. This cascaded effect becomes highly efficient when the pump pulse is pre-modulated with proper second and third order dispersion.
%This modulation leads to efficient multicycle THz pulse generation. %Our calculation shows that 1.6 times stronger THz conversion efficiency than the THz generation without considering the Pockels effect when pump pulse is highly chirped.
We also report the first demonstration of multicycle THz pulse generation at 14 THz from LT and 14.6 THz from $\beta$-BBO crystals. This new type of narrowband phase matching scheme is universal and can be applied to many nonlinear materials with potential to provide robust, efficient, and tabletop multicycle THz sources.
%with maximum output energy of 92.6 nJ with corresponding focused THz peak electric field of 0.4 MV/cm in the LN crystal, and this source would be useful for various applications.

\section{Appendix}
\subsection{Dispersion control in a dual grating compressor}
%A pair of diffraction gratings are commonly used for pulse compression in a chirped pulse amplification (CPA) laser system.
In this experiment, the laser pulse duration is controlled by adjusting the separation between a pair of diffraction gratings in the pulse compressor. %An analytical description of the dispersion are readily available for the dual grating compressor.
Right after the compressor, the second-order and third-order spectral phases of the laser pulse are given by \cite{druon2008simple}

\begin{equation}
\phi^{(2)}=-\frac{L\lambda^{3}}{\pi c^{2}d^{2}}\left[ 1 - \left( \lambda/d - \sin\theta_{in} \right)^{2} \right]^{-\frac{3}{2}},
\end{equation}
\begin{equation}
\phi^{(3)}=-\phi^{(2)}\frac{3\lambda}{2\pi c}\left[ \frac{1+ \sin\theta_{in}\left( \lambda/d - \sin\theta_{in} \right)}{1-\left( \lambda/d - \sin\theta_{in} \right)^{2}} \right],
\end{equation}
where \textit{L} is the perpendicular distance between the two parallel gratings, $\theta_{in}$ is the incident laser angle, and $d$ is the grating groove spacing. In our compressor system, we use 1,500 grooves/mm gratings at $\theta_{in} = 58.4^{\circ}$ around a central wavelength of $\lambda$ = 800 nm. The second-order $\phi^{(2)}$ and third-order $\phi^{(3)}$ are generally referred to GDD and TOD, respectively. %The grating separation \textit{L} is varied from 264 mm to 266 mm to vary GDD and TOD. 
Figure 10 plots how GDD and TOD vary as a function of the grating separation $L$ in our pulse compressor.

\begin{figure}[h!]
\centering\includegraphics[width=\textwidth,height=\textheight,keepaspectratio]{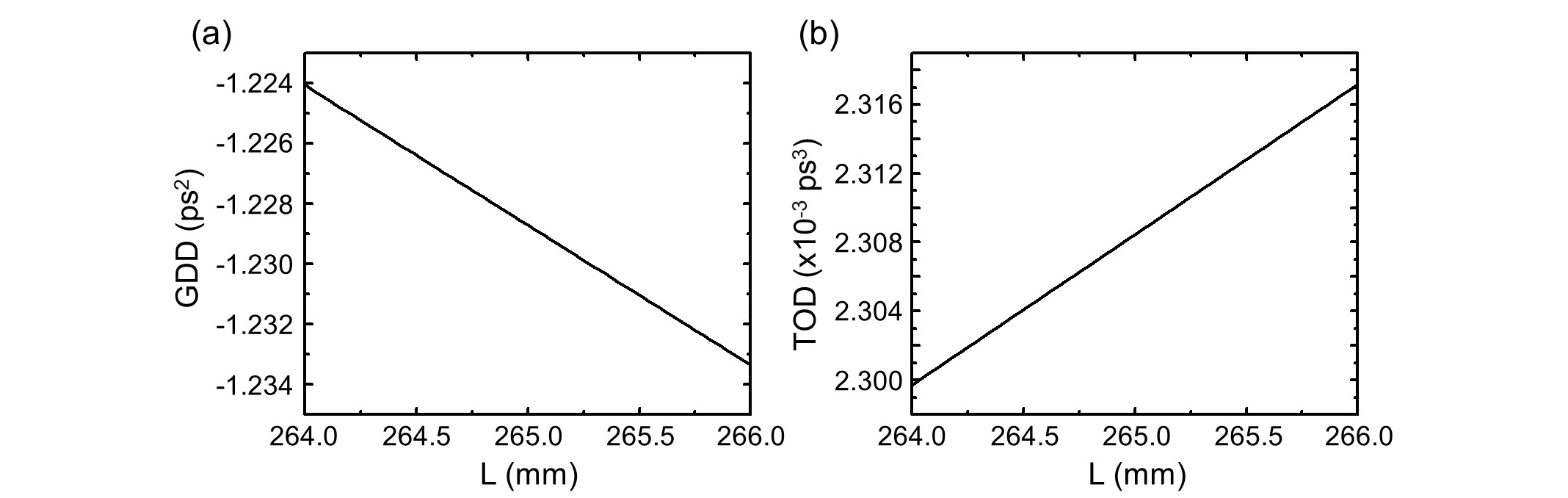}
\caption{(a) Second-order spectral phase (GDD) and (b) third-order spectral phase (TOD) introduced by a dual grating compressor as a function of the separation between the grating pair.}
\end{figure}

For simplicity, Eq. (8) can be rewritten as TOD = $\text{A}\cdot \text{GDD}$, where A is a constant that depends on both the laser wavelength and incident angle, i.e., A = A($\lambda,\theta_{in}$). In our case, it is estimated to A $\approx$ $-$2 fs, %1.88 fs
and the shortest pulse duration is achieved with GDD = $-$1.229 ps$^{2}$. Thus, a ``negative'' or ``positive'' chirp is defined when GDD becomes less or greater than $-$1.229 ps$^{2}$, respectively. %In our compressor system, GDD and TOD vary together from -4,700 fs$^{2}$ to 4,700 fs$^{2}$ and from -8,950 fs$^{3}$ to 8,950 fs$^{3}$, respectively.

\subsection{Calculation of $\epsilon (\omega)$ and $d_\text{eff}  (\omega)$ of lithium niobate}
The complex dielectric function $\epsilon (\omega)$ and nonlinear coefficient  $d_\text{eff}  (\omega)$ of congruent LN are given by  \cite{barker1967dielectric, sussman1970tunable}

\begin{equation}
\epsilon\left( \omega \right) = \epsilon_{\infty} + \sum_{j} \frac{S_{j}\omega_{j}^{2}}{\omega_{j}^{2} - \omega^{2} - i\omega\Gamma_{j}},
\end{equation}
\begin{equation}
d_{\text{eff}}\left( \omega \right) = d_{e} + \sum_{j} \frac{S_{j}\omega_{j}^{2}}{\omega_{j}^{2} - \omega^{2} - i\omega\Gamma_{j}}d_{Qj},
\end{equation}
where $S_{j}$, $\omega_{j}$, and $\Gamma_{j}$ are the oscillator strength, the resonance frequency, and the width of the $j$th transverse-phonon mode; $\epsilon_{\infty}$ = 4.6 is the frequency-independent bound electronic dielectric function; $d_{e}$ and $d_{Qj}$ are the electronic and ionic nonlinear coefficients, respectively. The real ($n$) and imaginary ($k$) parts of the square root of the complex dielectric function, $\sqrt{\epsilon\left( \omega \right)} = n + ik$, are related to the refractive index $n\left(\omega\right) = n$ and absorption coefficient $\alpha\left(\omega\right) = 2k\omega/c$. %The results are restricted to interactions in which all waves are polarized parallel to the extraordinary axis of LN crystal to be consistent with our experimental condition. 
All parameter values necessary to calculate Eqs. (9) and (10) are listed in table 1.

\begin{table}[h!]
\caption{Properties of transverse modes in the extraordinary axis of LN. Additional modes at 3.9 THz and 20.76 THz are included to calculate the dielectric function.%since group theory predicts only four transverse modes.
}
\centering 
\setlength{\tabcolsep}{10pt} % Default value: 6pt
\small						%Added to reduce font size
\begin{tabular}{|c c c c c|} 
\hline\hline %inserts double horizontal lines
$\omega_{j}/2\pi\left(\text{THz}\right)$ & $S_{j}$ & $\Gamma_{j}/2\pi \left(\text{THz}\right)$ & $d_{Qj}\left(\text{pm/V} \right)$ & \\ [0.5ex] 
\hline 
3.9 & 5.7 & 0.69 & - & \cite{schall1999far} \\ 
7.44 & 16 & 0.63 & 7.66 & \cite{barker1967dielectric,sussman1970tunable}\\
8.22 & 1 & 0.42 & -14.68 & \cite{barker1967dielectric,sussman1970tunable}\\
9.21 & 0.16 & 0.75 & -17.87 & \cite{barker1967dielectric,sussman1970tunable}\\
18.84 & 2.55 & 1.02 & 11.49 & \cite{barker1967dielectric,sussman1970tunable}\\
20.76 & 0.13 & 1.47 & - & \cite{barker1967dielectric}\\
\multicolumn{5}{|c|}{$\epsilon_{\infty}$=4.6 \cite{barker1967dielectric} \hspace{1cm} $d_{e}$=33.7 pm/V \cite{sussman1970tunable}}\\ [0.5ex] 
\hline 
\end{tabular}
\label{table:nonlin} 
\end{table}

\subsection{TOD effects on multicycle THz pulse generation}
To investigate the TOD effects, we repeated the simulation with various pump TOD values. In our experiment and simulation, the pump TOD varies with GDD as TOD = TOD$_g$ + TOD$_i$ = $-2(\text{fs})\cdot\text{GDD}$ + TOD$_i$. Figure 11 shows 2-D plots of narrowband THz energy (color scale) obtained with TOD$_i$ of $-$4000, $- $2000, 0, 2000, and 4000 fs$^{3}$. The pump fluence is set to 13.6 mJ/cm$^{2}$. For comparison, Fig. 8(a) is obtained with TOD$_i$ = 3,800 fs$^3$. 
%The simulated multi-cycle THz pulse energies are plotted in Fig. 11.

As shown in Fig. 11, the output THz energy strongly depends on both GDD and TOD. With  TOD$_i$ = 0 fs$^3$, efficient THz generation occurs near GDD = 0 fs$^2$ where the pump pulse duration remains relatively short. In this case, the energy rapidly increases but also quickly drops due to large THz absorption. %Here the optimal propagation length (or LN thickness) for THz generation is 40 $\mu$m, shorter than the effective length for maximal THz generation $L_{\text{eff}}$ = 160 $\mu$m obtained under ideal phase-matched optical rectification where no pump dispersion or distortion is assumed. Because of nonzero dispersion and $\chi^{(2)}$-induced nonlinearity, the optimal length becomes shorter than $L_{\text{eff}}$.
%slightly longer than $\alpha^{-1}$ because of a non-negligible contribution from the Pockels effect. %This could be treated as a simple optical rectification process without considering other nonlinear effects such as cross-phase modulation (XPM), self-phase modulation (SPM), multi-photon absorption and so on.
Due to normal material dispersion in LN (368 fs$^{2}$/mm at 800 nm), more negative GDD is necessary to keep the pump pulse duration short with increasing propagation distance (or LN thickness). This is why the color plot in Fig. 11(c) has a single stripe with a negative slope. %Also the optimal GDD is slightly negatively chirped due to normal dispersion in LN (368 fs$^{2}$/mm at 800 nm) in order to  efficient THz pulses from the exit plane of the LN crystal. 
With large TOD$_i$ values (either positive or negative), optimal THz generation occurs with relatively large GDD values (positive or negative) depending on the TOD$_i$ sign and the propagation distance $z$. In this regime, the pump intensity envelop can be pre-modulated by a proper combination of GDD and TOD, and the modulation can be resonantly amplified through the cascaded Pockels effect.

\begin{figure}[t!]
\centering\includegraphics[width=\textwidth,height=\textheight,keepaspectratio]{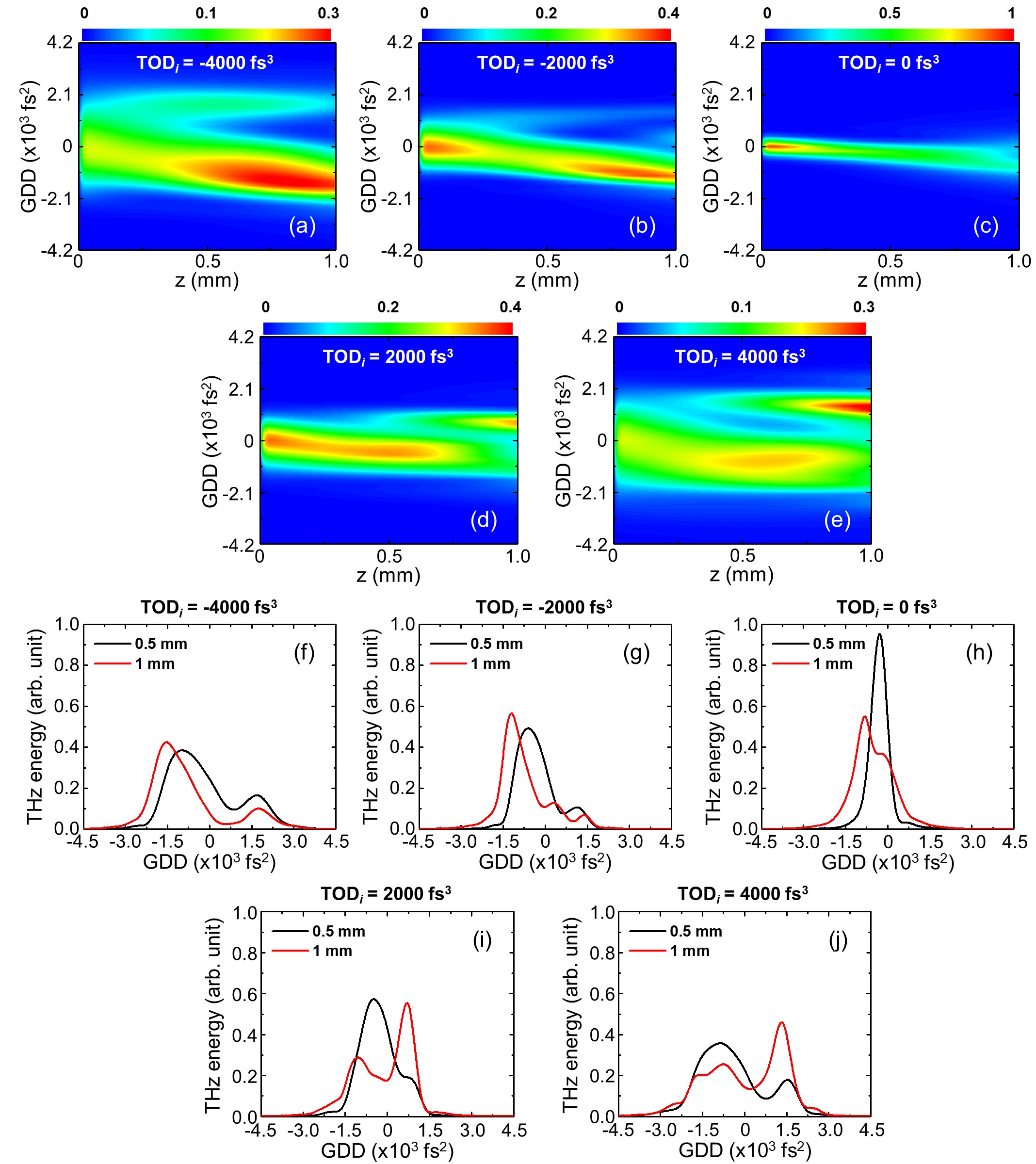}
\caption{(a-e) 2-D plots of multicycle THz energy (color scale) as a function of the propagation direction $z$ (horizontal) and initial pump GDD (vertical) with the initial (residual) TOD$_{i}$ of -4000, -2000, 0, 2000, and 4000 fs$^{3}$. (f-g) Line-outs at $z$ = 0.5 mm (black lines) and 1 mm (red lines).}
\end{figure}

\newpage

\section*{Funding}
Air Force Office of Scientific Research (FA9550- 25116-1-0163); Office of Naval Research (N00014-17-1-2705)

%\section*{Acknowledgments}
%Acknowledgments, if included, should appear at the end of the document. The section title should not be numbered.

\bibliographystyle{unsrt}
\bibliography{references}

\begin{thebibliography}{10}

\bibitem{nanni2015terahertz}
Emilio~A Nanni, Wenqian~R Huang, Kyung-Han Hong, Koustuban Ravi, Arya Fallahi,
  Gustavo Moriena, RJ~Dwayne Miller, and Franz~X K{\"a}rtner.
\newblock Terahertz-driven linear electron acceleration.
\newblock {\em Nature Communications}, 6:8486, 2015.

\bibitem{palfalvi2014evanescent}
L~P{\'a}lfalvi, JA~F{\"u}l{\"o}p, Gy~T{\'o}th, and J~Hebling.
\newblock Evanescent-wave proton postaccelerator driven by intense {TH}z pulse.
\newblock {\em Physical Review Special Topics-Accelerators and Beams},
  17(3):031301, 2014.

\bibitem{kampfrath2013resonant}
Tobias Kampfrath, Koichiro Tanaka, and Keith~A Nelson.
\newblock Resonant and nonresonant control over matter and light by intense
  terahertz transients.
\newblock {\em Nature Photonics}, 7(9):680, 2013.

\bibitem{schubert2014sub}
Olaf Schubert, Matthias Hohenleutner, Fabian Langer, Benedikt Urbanek, C~Lange,
  U~Huttner, D~Golde, T~Meier, M~Kira, Stephan~W Koch, et~al.
\newblock Sub-cycle control of terahertz high-harmonic generation by dynamical
  {Bloch} oscillations.
\newblock {\em Nature Photonics}, 8(2):119--123, 2014.

\bibitem{nicoletti2016nonlinear}
Daniele Nicoletti and Andrea Cavalleri.
\newblock Nonlinear light--matter interaction at terahertz frequencies.
\newblock {\em Advances in Optics and Photonics}, 8(3):401--464, 2016.

\bibitem{lee2009principles}
Yun-Shik Lee.
\newblock {\em Principles of Terahertz Science and Technology}, volume 170.
\newblock Springer Science \& Business Media, 2009.

\bibitem{hebling2008generation}
J{\'a}nos Hebling, Ka-Lo Yeh, Matthias~C Hoffmann, Bal{\'a}zs Bartal, and
  Keith~A Nelson.
\newblock Generation of high-power terahertz pulses by tilted-pulse-front
  excitation and their application possibilities.
\newblock {\em Journal of the Optical Society of America B}, 25(7):B6--B19,
  2008.

\bibitem{palik1997lithium}
Edward~D Palik.
\newblock Lithium niobate ({L}i{N}b{O}$_3$).
\newblock In {\em Handbook of Optical Constants of Solids}, pages 695--702.
  Elsevier, 1997.

\bibitem{fulop2014efficient}
J{\'o}zsef~A F{\"u}l{\"o}p, Zolt{\'a}n Ollmann, Cs~Lombosi, Christoph Skrobol,
  Sandro Klingebiel, L{\'a}szl{\'o} P{\'a}lfalvi, Ferenc Krausz, Stefan Karsch,
  and J{\'a}nos Hebling.
\newblock Efficient generation of {TH}z pulses with 0.4 m{J} energy.
\newblock {\em Optics Express}, 22(17):20155--20163, 2014.

\bibitem{ravi2014limitations}
Koustuban Ravi, W~Ronny Huang, Sergio Carbajo, Xiaojun Wu, and Franz
  K{\"a}rtner.
\newblock Limitations to {TH}z generation by optical rectification using tilted
  pulse fronts.
\newblock {\em Optics Express}, 22(17):20239--20251, 2014.

\bibitem{wu2018highly}
Xiao-Jun Wu, Jing-Long Ma, Bao-Long Zhang, Shu-Su Chai, Zhao-Ji Fang, Chen-Yi
  Xia, De-Yin Kong, Jin-Guang Wang, Hao Liu, Chang-Qing Zhu, et~al.
\newblock Highly efficient generation of 0.2 m{J} terahertz pulses in lithium
  niobate at room temperature with sub-50 fs chirped {T}i:sapphire laser
  pulses.
\newblock {\em Optics Express}, 26(6):7107--7116, 2018.

\bibitem{wong2013compact}
Liang~Jie Wong, Arya Fallahi, and Franz~X K{\"a}rtner.
\newblock Compact electron acceleration and bunch compression in {TH}z
  waveguides.
\newblock {\em Optics Express}, 21(8):9792--9806, 2013.

\bibitem{kartner2016axsis}
FX~K{\"a}rtner, Frederike Ahr, A-L Calendron, H~{\c{C}}ankaya, S~Carbajo,
  G~Chang, G~Cirmi, K~D{\"o}rner, U~Dorda, A~Fallahi, et~al.
\newblock {AXSIS}: Exploring the frontiers in attosecond {X}-ray science,
  imaging and spectroscopy.
\newblock {\em Nuclear Instruments and Methods in Physics Research Section A:
  Accelerators, Spectrometers, Detectors and Associated Equipment}, 829:24--29,
  2016.

\bibitem{lee2000generation}
Y-S Lee, T~Meade, V~Perlin, H~Winful, TB~Norris, and A~Galvanauskas.
\newblock Generation of narrow-band terahertz radiation via optical
  rectification of femtosecond pulses in periodically poled lithium niobate.
\newblock {\em Applied Physics Letters}, 76(18):2505--2507, 2000.

\bibitem{l2007generation1}
JA~L’huillier, G~Torosyan, M~Theuer, Yu~Avetisyan, and R~Beigang.
\newblock Generation of {TH}z radiation using bulk, periodically and
  aperiodically poled lithium niobate--{P}art 1: Theory.
\newblock {\em Applied Physics B}, 86(2):185--196, 2007.

\bibitem{l2007generation2}
JA~L’huillier, G~Torosyan, M~Theuer, C~Rau, Yu~Avetisyan, and R~Beigang.
\newblock Generation of {TH}z radiation using bulk, periodically and
  aperiodically poled lithium niobate--{P}art 2: Experiments.
\newblock {\em Applied Physics B}, 86(2):197--208, 2007.

\bibitem{carbajo2015efficient}
Sergio Carbajo, Jan Schulte, Xiaojun Wu, Koustuban Ravi, Damian~N Schimpf, and
  Franz~X K{\"a}rtner.
\newblock Efficient narrowband terahertz generation in cryogenically cooled
  periodically poled lithium niobate.
\newblock {\em Optics Letters}, 40(24):5762--5765, 2015.

\bibitem{jolly2019spectral}
Spencer~W Jolly, Nicholas~H Matlis, Frederike Ahr, Vincent Leroux, Timo
  Eichner, Anne-Laure Calendron, Hideki Ishizuki, Takunori Taira, Franz~X
  K{\"a}rtner, and Andreas~R Maier.
\newblock Spectral phase control of interfering chirped pulses for high-energy
  narrowband terahertz generation.
\newblock {\em Nature Communications}, 10(1):2591, 2019.

\bibitem{chen2011generation}
Zhao Chen, Xibin Zhou, Christopher~A Werley, and Keith~A Nelson.
\newblock Generation of high power tunable multicycle teraherz pulses.
\newblock {\em Applied Physics Letters}, 99(7):071102, 2011.

\bibitem{stepanov2004generation}
AG~Stepanov, J~Hebling, and Juergen Kuhl.
\newblock Generation, tuning, and shaping of narrow-band, picosecond {TH}z
  pulses by two-beam excitation.
\newblock {\em Optics Express}, 12(19):4650--4658, 2004.

\bibitem{lin2009generation}
Kung-Hsuan Lin, Christopher~A Werley, and Keith~A Nelson.
\newblock Generation of multicycle terahertz phonon-polariton waves in a planar
  waveguide by tilted optical pulse fronts.
\newblock {\em Applied Physics Letters}, 95(10):232, 2009.

\bibitem{cirmi2017cascaded}
Giovanni Cirmi, Michael Hemmer, Koustuban Ravi, Fabian Reichert, Luis~E Zapata,
  Anne-Laure Calendron, H{\"u}seyin {\c{C}}ankaya, Frederike Ahr, Oliver~D
  M{\"u}cke, Nicholas~H Matlis, et~al.
\newblock Cascaded second-order processes for the efficient generation of
  narrowband terahertz radiation.
\newblock {\em Journal of Physics B: Atomic, Molecular and Optical Physics},
  50(4):044002, 2017.

\bibitem{jang2019hidden}
Dogeun Jang, Yung~Jun Yoo, and Ki-Yong Kim.
\newblock Hidden phase-matched narrowband {TH}z generation via optical
  rectification in lithium niobate.
\newblock In {\em CLEO: Science and Innovations}, pages STh3F--4. Optical
  Society of America, 2019.

\bibitem{jang2020generation}
Dogeun Jang, Jae~Hee Sung, Seong~Ku Lee, Chul Kang, and Ki-Yong Kim.
\newblock Generation of 0.7 m{J} multicycle 15 {TH}z radiation by phase-matched
  optical rectification in lithium niobate.
\newblock {\em arXiv preprint arXiv:2003.11715}, 2020.

\bibitem{lin2012efficient}
Xiaomu Lin, Lei Wang, and Yujie~J Ding.
\newblock Efficient generation of far-infrared radiation in the vicinity of
  polariton resonance of lithium niobate.
\newblock {\em Optics Letters}, 37(17):3687--3689, 2012.

\bibitem{chen2010frequency}
Ching-Wei Chen, Yen-Cheng Lin, Chia-Hua Chang, Peichen Yu, Jia-Min Shieh, and
  Ci-Ling Pan.
\newblock Frequency-dependent complex conductivities and dielectric responses
  of indium tin oxide thin films from the visible to the far-infrared.
\newblock {\em IEEE Journal of Quantum Electronics}, 46(12):1746--1754, 2010.

\bibitem{yoo2016generation}
Yung-Jun Yoo, Donghoon Kuk, Zheqiang Zhong, and Ki-Yong Kim.
\newblock Generation and characterization of strong terahertz fields from k{H}z
  laser filamentation.
\newblock {\em IEEE Journal of Selected Topics in Quantum Electronics},
  23(4):1--7, 2016.

\bibitem{jang2019spectral}
Dogeun Jang, Malik Kimbrue, Yung-Jun Yoo, and Ki-Yong Kim.
\newblock Spectral characterization of a microbolometer focal plane array at
  terahertz frequencies.
\newblock {\em IEEE Transactions on Terahertz Science and Technology},
  9(2):150--154, 2019.

\bibitem{jang2019scalable}
Dogeun Jang, Chul Kang, Seong~Ku Lee, Jae~Hee Sung, Chul-Sik Kee, Seung~Woo
  Kang, and Ki-Yong Kim.
\newblock Scalable terahertz generation by large-area optical rectification at
  80 {TW} laser power.
\newblock {\em Optics Letters}, 44(22):5634--5637, 2019.

\bibitem{schall1999far}
M~Schall, H~Helm, and SR~Keiding.
\newblock Far infrared properties of electro-optic crystals measured by {TH}z
  time-domain spectroscopy.
\newblock {\em Journal of Infrared, Millimeter, and Terahertz Waves},
  20(4):595--604, 1999.

\bibitem{zhong2015optimization}
Sen-Cheng Zhong, Zhao-Hui Zhai, Jiang Li, Li-Guo Zhu, Jun Li, Kun Meng, Qiao
  Liu, Liang-Hui Du, Jian-Heng Zhao, and Ze-Ren Li.
\newblock Optimization of terahertz generation from {L}i{N}b{O}$_3$ under
  intense laser excitation with the effect of three-photon absorption.
\newblock {\em Optics Express}, 23(24):31313--31323, 2015.

\bibitem{fulop2010design}
JA~F{\"u}l{\"o}p, L~P{\'a}lfalvi, G~Alm{\'a}si, and J~Hebling.
\newblock Design of high-energy terahertz sources based on optical
  rectification.
\newblock {\em Optics Express}, 18(12):12311--12327, 2010.

\bibitem{gervais1997lithium}
Fran{\c{c}}ois Gervais and Vicente Fonseca.
\newblock Lithium tantalate ({L}i{T}a{O}$_3$).
\newblock In {\em Handbook of Optical Constants of Solids}, pages 777--805.
  Elsevier, 1997.

\bibitem{valverde2017multi}
David~A Valverde-Ch{\'a}vez and David~G Cooke.
\newblock Multi-cycle terahertz emission from $\beta$-barium borate.
\newblock {\em Journal of Infrared, Millimeter, and Terahertz Waves},
  38(1):96--103, 2017.

\bibitem{ding2011efficient}
Yujie~J Ding.
\newblock Efficient generation of far-infrared radiation from a periodically
  poled {L}i{N}b{O}$_3$ waveguide based on surface-emitting geometry.
\newblock {\em Journal of the Optical Society of America B}, 28(5):977--981,
  2011.

\bibitem{sussman1970tunable}
Stanley~Saul Sussman.
\newblock Tunable light scattering from transverse optical modes in lithium
  niobate.
\newblock Technical report, STANFORD UNIV CA MICROWAVE LAB, 1970.

\bibitem{barker1967dielectric}
AS~Barker~Jr and R~Loudon.
\newblock Dielectric properties and optical phonons in {L}i{N}b{O}$_3$.
\newblock {\em Physical Review}, 158(2):433, 1967.

\bibitem{hattori2007simulation}
Toshiaki Hattori and Kousuke Takeuchi.
\newblock Simulation study on cascaded terahertz pulse generation in
  electro-optic crystals.
\newblock {\em Optics Express}, 15(13):8076--8093, 2007.

\bibitem{shen2007nonlinear}
Y~Shen, T~Watanabe, DA~Arena, C-C Kao, JB~Murphy, TY~Tsang, XJ~Wang, and
  GL~Carr.
\newblock Nonlinear cross-phase modulation with intense single-cycle terahertz
  pulses.
\newblock {\em Physical Review Letters}, 99(4):043901, 2007.

\bibitem{jewariya2009enhancement}
Mukesh Jewariya, Masaya Nagai, and Koichiro Tanaka.
\newblock Enhancement of terahertz wave generation by cascaded $\chi^{(2)}$
  processes in {L}i{N}b{O}$_3$.
\newblock {\em Journal of the Optical Society of America B}, 26(9):A101--A106,
  2009.

\bibitem{husakou2001supercontinuum}
AV~Husakou and J~Herrmann.
\newblock Supercontinuum generation of higher-order solitons by fission in
  photonic crystal fibers.
\newblock {\em Physical Review Letters}, 87(20):203901, 2001.

\bibitem{desalvo1996infrared}
Richard DeSalvo, Ali~A Said, David~J Hagan, Eric~W Van~Stryland, and Mansoor
  Sheik-Bahae.
\newblock Infrared to ultraviolet measurements of two-photon absorption and
  $n_2$ in wide bandgap solids.
\newblock {\em IEEE Journal of Quantum Electronics}, 32(8):1324--1333, 1996.

\bibitem{druon2008simple}
Fr{\'e}d{\'e}ric Druon, Marc Hanna, Ga{\"e}lle Lucas-Leclin, Yoann Zaouter,
  Dimitris Papadopoulos, and Patrick Georges.
\newblock Simple and general method to calculate the dispersion properties of
  complex and aberrated stretchers-compressors.
\newblock {\em Journal of the Optical Society of America B}, 25(5):754--762,
  2008.

\end{thebibliography}
\end{document}